\documentclass[
prc,preprint,nofootinbib,preprintnumbers,
superscriptaddress,tightenlines
]{revtex4-1}

\usepackage{amsmath}    
\usepackage{graphicx}   
\usepackage{subfigure}
\usepackage{color}
\usepackage[colorlinks=true,linkcolor=blue,citecolor=blue, urlcolor=blue]{hyperref}   
\usepackage{bm} 
\usepackage{ulem}
\usepackage{soul}
\usepackage{graphicx}
\usepackage{amsmath,amssymb}
\usepackage{soul}
\usepackage{textcomp}
\usepackage{hyperref}
\usepackage{subfigure}
\usepackage[toc,page]{appendix}

\def\crs{{\rm cr}}
\def\Cp{{C_p}}

\newcommand{\blue}[1]{\textcolor{blue}{#1}}

\def \pd {\partial}

\def\st{\begin{equation}}
\def\stp{\end{equation}}
\def\bg{\begin{eqnarray}}
\def\nd{\end{eqnarray}}
\def\Eq#1{Eq.~(\ref{#1})}
\def\Eqs#1{Eqs.~(\ref{#1})}
\def\eq#1{(\ref{#1})}
\def\app#1{Appendix~\ref{#1}}
\def\Fig#1{Fig.~\ref{#1}}

\def\Sect#1{Sect.~\ref{#1}}
\def\Ref#1{Ref.~\cite{#1}}

\def\llangle{\left\langle}
\def\rrangle{\right\rangle}
\def \bes {\begin{subequations}}
\def \ees {\end{subequations}}

\def \S{{\mathcal S}}

\def\nh{{\hat n}}
\def\E{{\mathcal E}}

\def\kk{\bm k}

\def\eis{{\epsilon}}
\def\chiis{{\chi_{\rm is}}}
\def\tcis{T^{\rm is}_c}
\def\ris{{r}}
\def\his{{h}}
\def\Gi{\mathcal G_{\rm is}}
\def\GQCD{\mathcal G}
\def\SQCD{\mathcal S}
\def\sh{\hat s}

\def\Asp{\phantom{A}}
\def\Si{\mathcal S^{\rm is}}
\def\muh{\hat{\mu}}
\def\minv{\bar{M}}
\def\M{{\mathcal M}}
\def\is{{\rm is}}

\def\r{\bm r}

\def \a {\alpha}
\def \b {\beta}

\def \d {\delta}

\def \g {\gamma}
\def \k {{\bm k}}

\def \l {\lambda}

\def \G {\Gamma}

\def \vx {\bm{x}}

\def \vk {\bm{k}}

\def \<{\langle}
\def \>{\rangle}
\def \+{\dagger}
\def \({\left(}
\def \){\right)}
\def \[{\left[}
\def \]{\right]}
\def\Chi{{\mathcal X}_{\rm is}}
\def\tb{{\bar t}}

\def\kb{{k\xi}}

\def\kz{{\rm kz}}
\def\KZ{{\rm kz}}
\def\notequal{{\neq}}
\def\neql{{\rm neq}}
\def\tkz{t_\kz}
\def\sp{\phantom{a}}
\def\tcross{{t_{\rm cr}}}
\def\cross{{\rm cr}}

\newcommand{\dfluct}[1]{\delta{#1}}

\def\sec{Sect.}

\def \shat{{\hat{s}}}

\def \shat{{\hat{s}}}

\def \ellmax{{\ell_{\rm max}}}
\def \xx{{\bm x}}
\def \yy{{\bm y}}
\def \tKZ {t_{\KZ}}
\def \lKZ {\ell_{\KZ}}

\def \Nn{N}
\def \Neq{N_0}

\def \Kchi{K_\chi}
\def \Nss {{N}^{\shat\shat}} 
\def \Nnn {{N}^{\hat n\hat n}} 
\def \Nssbar {\bar{N}^{\shat\shat}} 

\begin{document}


\title{Transits of the QCD Critical Point }

\author{Yukinao Akamatsu}
\email[]{akamatsu@kern.phys.sci.osaka-u.ac.jp}
\affiliation{Department of Physics, Osaka University, Toyonaka, Osaka 560-0043, Japan}
\author{Derek Teaney}
\email[]{derek.teaney@stonybrook.edu}
\affiliation{Department of Physics and Astronomy, Stony Brook University, Stony Brook, New York 11794, USA}
\author{Fanglida Yan}
\email[]{yan.fanglida@stonybrook.edu}
\affiliation{Department of Physics and Astronomy, Stony Brook University, Stony Brook, New York 11794, USA}
\author{Yi Yin}
\email[]{yiyin3@mit.edu}
\affiliation{Center for Theoretical Physics, Massachusetts Institute of Technology, Cambridge, Massachusetts 02139, USA }

\preprint{MIT-CTP/5042}
\date{\today}

   \begin{abstract}
   We analyze the evolution of hydrodynamic fluctuations
in a heavy ion collision as the system passes close to the QCD critical
      point. We introduce two small dimensionless parameters $\lambda$ and
      $\Delta_s$ to characterize the evolution. $\lambda$ compares the microscopic relaxation time (away from
      the critical point) to the expansion rate $\lambda \equiv \tau_0/\tau_Q$, and $\Delta_s$ compares the  baryon to entropy ratio, $n/s$, to its
      critical value, $\Delta_s\equiv (n/s - n_c/s_c)/(n_c/s_c)$. We determine
      how the evolution of critical hydrodynamic fluctuations depends parametrically 
      on $\lambda$ and $\Delta_s$. Finally, we use this parametric reasoning to estimate the critical fluctuations and correlation length for a  heavy ion collision, and to give guidance to the experimental search for the QCD critical point.
   \end{abstract}


\maketitle
\section{Introduction}
\label{sec:intro}

\subsection{Overview and goals}

%
%
%
The conjectured QCD critical point is a landmark point in the QCD phase diagram. 
This is the end point of a line of first-order phase transitions, which separates the Quark-Gluon Plasma (QGP) phase from  hadronic matter. 
Due to the sign problem at finite baryon chemical potential, lattice QCD simulations have yet to confirm the existence of a critical point~\cite{Ding:2015ona}.
Nevertheless, the conjectured point in the phase diagram is theoretically well motivated, and  has been found in various effective field theory models, see Refs.~\cite{Stephanov:2004wx,Stephanov:2007fk,Fukushima:2010bq} for reviews. 
An  intense experimental effort is underway to
locate and to characterize the critical point through a beam energy scan (BES) of heavy ion collisions at the Relativistic Heavy Ion Collider (RHIC)~\cite{STAR-wp,Luo:2017faz}.  

%
%
%
The experimental search for the QCD critical point will focus on fluctuations.
The existence of a critical point in a heavy ion collision should lead to large correlations and enhanced fluctuations of conserved densities~\cite{Stephanov:1998dy, Stephanov:1999zu}. 
These enhanced fluctuations should manifest themselves through the multiplicity fluctuations of the produced hadrons.
However,  the
systems created in these nuclear collisions are rapidly expanding, and consequently thermodynamic fluctuations will not be fully equilibrated.
In particular, 
it has been  demonstrated previously
that  
due to the expansion of the fireball and the physics of critical
slowing down,
the critical fluctuations can differ significantly  from their equilibrium
expectation~\cite{Berdnikov:1999ph, Mukherjee:2015swa}. 
Further, in any real experiment the system will not 
pass directly through the critical point,  and this again will limit the size of the critical fluctuations.   

%
To quantify how the expansion of the system and missing the critical point will
tame the critical fluctuations  we will introduce two small
parameters,
$\lambda$ and $\Delta_s$,
which characterize the evolution of the fireball:
\begin{align}
   \lambda \equiv& \frac{\tau_0}{\tau_Q} \, ,  \\
 \Delta_s \equiv& \frac{n_c}{s_c} \left(\frac{s}{n} - \frac{s_c}{n_c} \right)  \, .
\end{align}
The first parameter $\lambda$ is the product of the microscopic relaxation time away from the critical point $\tau_0$
and the expansion rate $1/\tau_Q$ (more precise definitions of $\tau_0$ and $\tau_Q$ are given below). The second parameter
$\Delta_s$ quantifies the deviation of the baryon number to entropy ratio $n/s$ from its critical value $n_c/s_c$ during the  adiabatic expansion of the system. A primary goal of the current
study is to determine  how the magnitude of the critical
fluctuations depends parametrically on these two small parameters.

%
%
In perfect equilibrium, the hydrodynamic fluctuations in the energy density (for example) 
are given by the text book thermodynamic formula
\st
\label{eq:edensityflucts}
\left. \llangle  \delta e(t,\xx)  \, \delta e (t,\yy) \rrangle \right|_{\rm equilibrium} = T^2 C_v \, \delta^{(3)}(\xx -\yy) \, ,
\stp
where $C_v$ is the specific heat at constant volume.
In Fourier space this says that all wavenumbers have equal amplitude
\st
\left. \llangle \delta e(t,\k) \, \delta e(t,-\k') \rrangle\right|_{\rm equilibrium} = T^2 C_v \, (2\pi)^3 \delta^{(3)}(\k - \k') \, .
\stp
However, for an expanding system, even away from the critical point, the distribution of fluctuations  
will not follow this equilibrium form, since long wavelengths of conserved quantities take a long time to relax to
equilibrium.  
The second goal of this paper is to determine the wavelength which characterizes the  
enhanced specific heats near the critical point, and to specify how this wavelength depends on $\lambda$ and $\Delta_s$.  

Away from the critical point, there is a length scale $\ell_{\rm max}$ where modes with wavelength longer than $\ell_{\rm max}$ 
fall out of equilibrium and reflect the expansion history rather than the equilibrium specific heat~\cite{Akamatsu:2016llw}. Indeed, the equilibration of
hydrodynamic fluctuations is a diffusive process. The diffusion coefficient
away from the
critical point is of order $D_0 \sim \ell_0^2 /\tau_0$ where $\tau_0$ is the relaxation time introduced above, 
and $\ell_0$ is a microscopic length.  
The maximum wavelength that can be equilibrated by diffusion over the total time time $\tau_Q$ is\footnote{In \Ref{Akamatsu:2016llw}  the length scale $\ell_{\rm max}$ is parametrized by the wavenumber $k_{*} \sim 1/\ell_{\rm max}$. }
\st
    \ell_{\rm max}^2 \sim  \ell_0^2 \left(\frac{\tau_Q}{\tau_0} \right) \, ,
\stp
or
\st
\label{lmax}
     \ell_{\rm max}  \sim \frac{\ell_0  }{\sqrt{\lambda} } \, .
\stp
There is insufficient time to equilibrate modes longer than $\ell_{\rm max}$,
and thus $\ell_{\rm max}$  provides  a robust upper cutoff on the 
size of critically correlated domains in the expanding fireball.

Near a critical point the diffusion coefficient is not a constant value $D_0$, but rapidly approaches zero. Thus the 
length scale characterizing critical domains is necessarily smaller than  $\ell_{\rm max}$.  
Modes with wavelength $\ell \ll \ell_{\rm max}$  
(but still longer than $\ell_0$) are equilibrated away from the critical point, but fall out of equilibrium as the
system approaches the critical point.
The emergent length scale, which arises from the competition between the 
expansion of the fireball and the diffusive equilibration of fluctuations, is known as the Kibble-Zurek length $\ell_{\KZ}$.  
The Kibble-Zurek length is the correlation length  at the time when the system falls out of equilibrium, and
characterizes both the magnitude and distribution of fluctuations
 in an evolving critical system~\cite{KIBBLE1980183,Zurek1985,Zurek:1996sj,PhysRevB.86.064304}.
The importance of Kibble-Zurek length (and time)  for the QCD critical point search  has been identified in Ref.~\cite{Mukherjee:2016kyu}.  As we  will see, the Kibble-Zurek length is of order
\st
  \ell_{\KZ}  \sim \frac{\ell_0 }{\lambda^{0.18} } \, , 
\stp
leading to an interesting hierarchy of scales $ \ell_0 \ll \ell_{\KZ} \ll \ell_{\max}$.
Both $\ell_{\rm max}$ and $\ell_{\KZ}$ are estimated in the conclusions.

Beyond parametric estimates, we will determine the time evolution of hydrodynamic correlators (such as \Eq{eq:edensityflucts})
by evolving stochastic hydrodynamics for an expanding fluid
in the vicinity of the QCD critical point.  Specifically, following
\Ref{Akamatsu:2016llw} (see
also Ref.~\cite{andreev1978corrections,Stephanov:2017ghc}). 
we will write down and solve a set hydro-kinetic equations governing the evolution of hydrodynamic two point functions.
The hydro-kinetic approach reformulates stochastic hydrodynamics as non-fluctuating hydrodynamics (describing a long wavelength background) coupled to a set of deterministic kinetic equations describing the phase
space distribution of short wavelength thermodynamic fluctuations, 
see also Refs.~\cite{Stephanov:2017ghc,Pratt:2017oyf} for related developments. 
The hydro-kinetic formulation successfully describes  non-trivial effects
such as the hydrodynamic tails~\cite{Akamatsu:2016llw} and the renormalization of bulk
viscosity~\cite{Akamatsu:2017rdu}, both of which are a consequence of the non-equilibrium evolution of thermodynamic fluctuations.
We first extend this approach to a system with non-zero net baryon density, and then implement critical fluctuations as implied by the critical universality. 
We  show how characteristic length scale $\ell_{\KZ}$ emerges from the hydro-kinetic equations for an expanding fireball.
See also Refs.~\cite{Kapusta:2012zb,Plumberg:2017tvu,Sakaida:2017rtj,Nahrgang:2017hkh} for previous studies of critical fluctuations based on stochastic hydrodynamics.

\subsection{Setup and outline}
\label{sec:trajectories}

\subsubsection{Setup}
Consider the hydrodynamic evolution of a single fluid cell of QCD matter passing close to the critical point. 
In the rest frame of the material, the entropy and baryon number densities follow the equations 
of ideal  hydrodynamics
\begin{subequations}
   \label{eq:expandingbox}
\begin{align}
   \partial_\tau s =& - s \,  \nabla \cdot u \, ,\\
   \partial_\tau n =& - n \, \nabla \cdot u \, ,
\end{align}
\end{subequations}
where $\tau$ is the proper time of the fluid cell and $\nabla \cdot u = \partial_{\mu} u^{\mu}$ is the expansion 
scalar.
Since the system is close to the critical point only for a short period of 
time we may treat the expansion scalar as a constant, $\partial_{\mu} u^{\mu} \equiv 1/\tau_Q$. 
Indeed, $\tau_Q$  is of order the system's lifetime, while  the time scales for the critical dynamics $t_{\KZ}$ and $t_{\cross}$ will be parametrically smaller than $\tau_Q$ justifying this approximation a-posteriori.

The entropy per baryon $s/n$ in \Eq{eq:expandingbox} is  constant in time.
We will refer to relative deviation of $s/n$  
from $s_c/n_c$ as the ``detuning" parameter, $\Delta_s$. 
Close to the critical point
\st
 \Delta_s \equiv \frac{n_c}{s_c} \left(\frac{s}{n} - \frac{s_c}{n_c} \right) \simeq \frac{\Delta s}{s_c} -  \frac{\Delta n}{n_c} \, , 
\stp
where $\Delta n$ 
notates the deviation from the critical value
\begin{align}
   \Delta n \equiv& n - n_c \,, 
\end{align}
with an analogous notation for other quantities (e.g. $\Delta \mu\equiv \mu - \mu_c$).
$\Delta_s$ is a dimensionless number and is small for a system passing close to the critical point.

There is a time $\tau_{1}$  where the baryon number reaches its critical value, $n_c$.  The
entropy at this time differs from its critical value  by $\Delta\bar{s}/{s_c} \simeq \Delta_s$.
For times close to $\tau_{1}$, we can integrate the equations of motion \Eq{eq:expandingbox} yielding
\begin{subequations}
   \label{eq:dsoversvstime}
\begin{align}
   \frac{ \Delta n(t)}{n_c} =&  - \frac{t}{\tau_Q} \, , \\
  \frac{\Delta s(t)}{s_c} =&  \Delta_s - \frac{t}{\tau_Q} \, , 
\end{align}
\end{subequations}
where we have defined $ t \equiv \tau - \tau_1$.
Thermodynamics relates the deviation in the (average) energy density from its critical value to these two quantities
\st
\label{eq:devstime}
  \Delta e  = T_c \Delta s +  \mu_c \Delta n  \,.
\stp

\begin{figure}
   \begin{center}
   \includegraphics[width=0.40\textwidth]{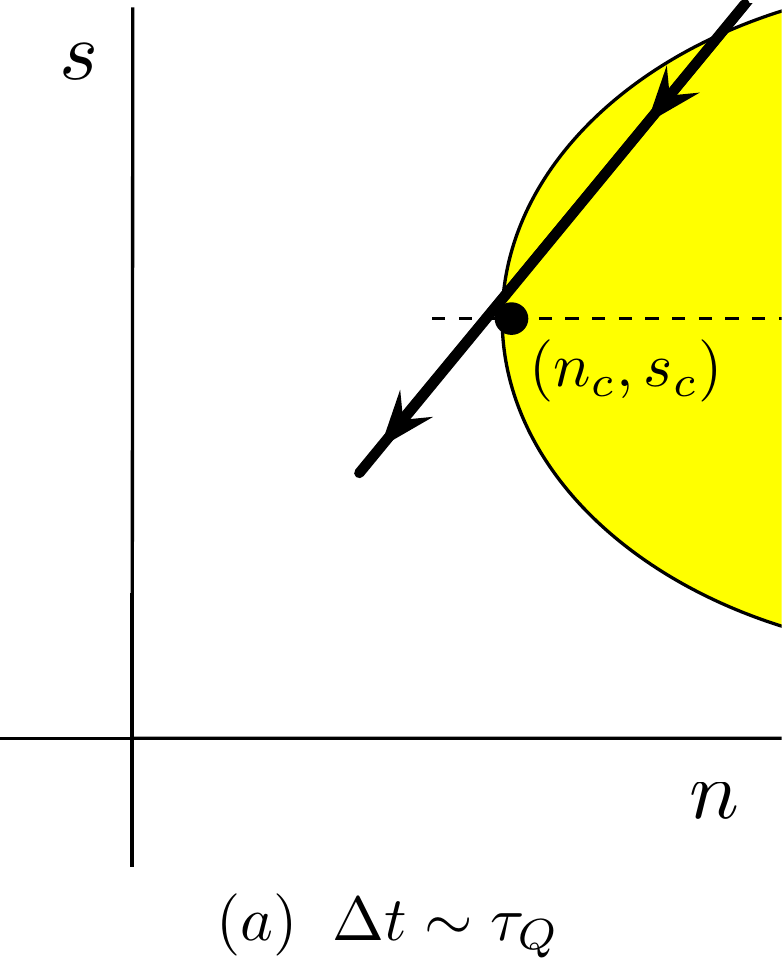}
      \hspace{0.1\textwidth}
   \includegraphics[width=0.40\textwidth]{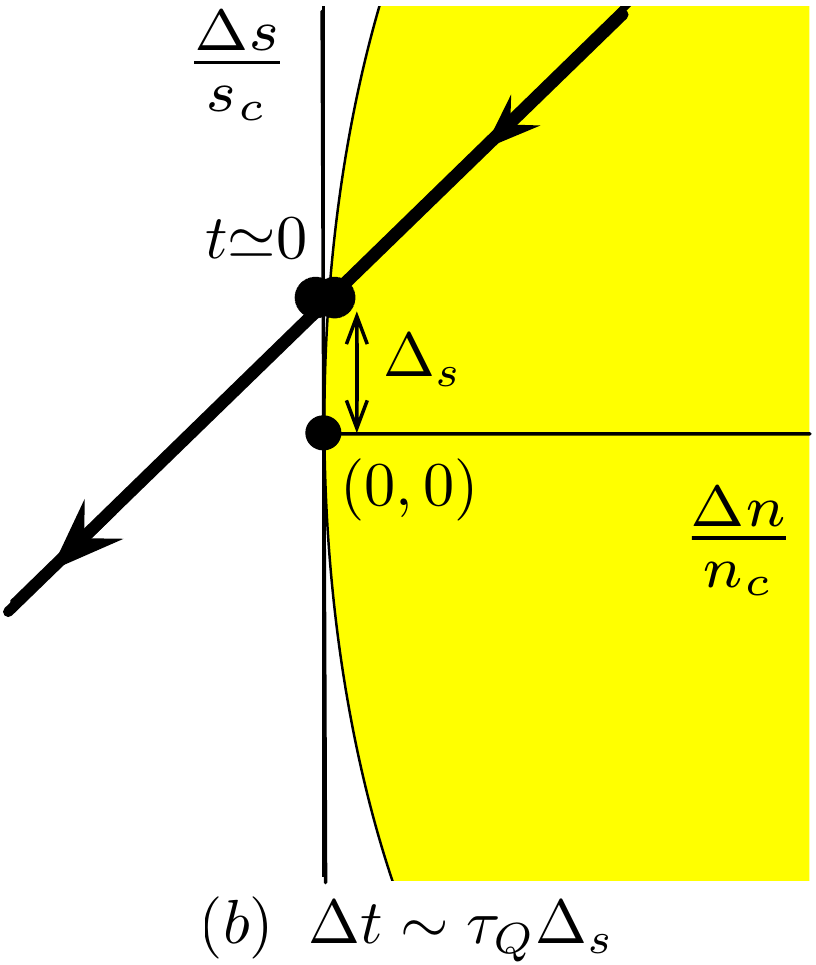} 
      \vskip 0.10in
   \includegraphics[width=0.40\textwidth]{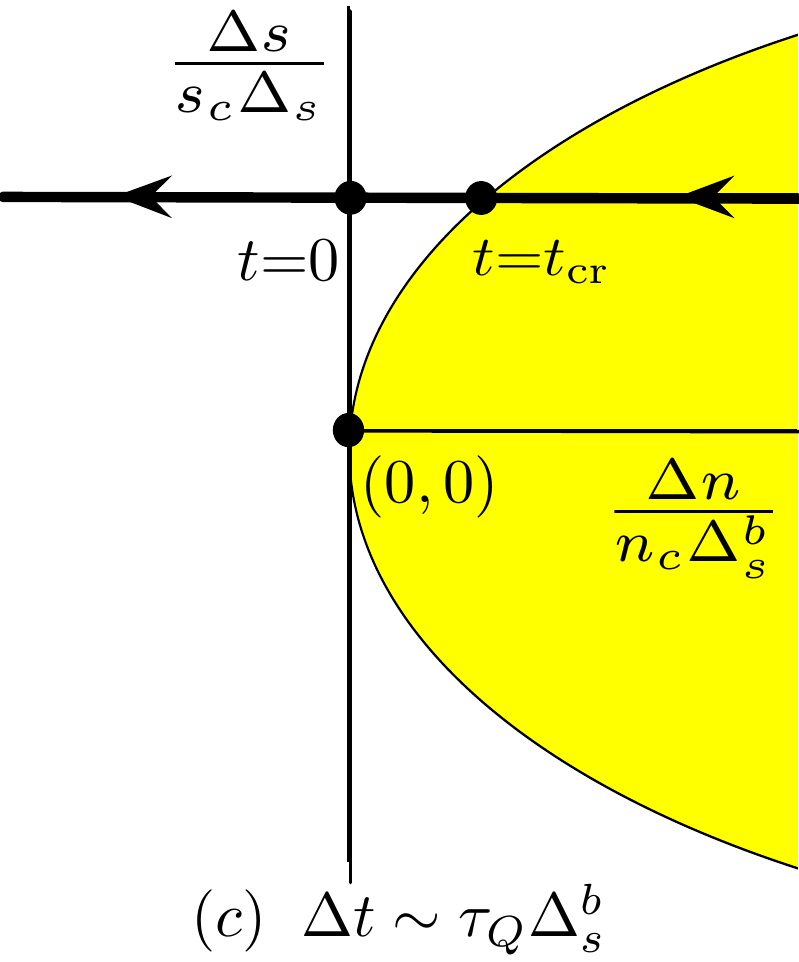}
      \caption{\label{trajectory}
      (a)  A schematic trajectory of a heavy ion collision passing close to the
      critical point. The duration of panel $(a)$ is of order $\Delta t \sim \tau_Q$.
      (b)  A magnification of the critical region in figure (a) by $\Delta_s$.
      The duration of panel $(b)$ is of order $\Delta t \sim \tau_Q \Delta_s$.
      In this regime the Ising magnetic field $h$ is negligibly small, and the susceptibilities
      scale as a power $\Delta n/n_c$.
      (c)  In this panel we have rescaled the $\Delta n/n_c$ and $\Delta s/s_c$ axes of (b) by $\Delta_s^b$
      and $\Delta_s$ respectively, with $b \equiv (1-\alpha)/\beta\simeq 2.7$. 
      The duration of panel (c) is of order $\Delta t \sim t_{\cross} \sim \tau_Q \Delta_s^b$.
      At the time $t_{\cross}$  the system leaves the coexistence region
      and the equilibrium correlation length reaches its maximal value (see \Eq{eq:tcross}).
      Only in panel (c) is the equation of state is a nontrivial function (i.e. beyond simple powers) of the scaling
      variable $z \propto r/h^{1/\beta\delta}$.}
   \end{center}
\end{figure}

In \Fig{trajectory}(a) we show a schematic picture of typical trajectory in the full QCD phase diagram, portrayed in the $(n,s)$-plane\footnote{
In this figure the coexistence line is shown as a flat line, which is a commonly used idealization~\cite{Nonaka:2004pg}.
This idealization is not essential to the parametric reasoning discussed in the text and illustrated in \Fig{trajectory}}.
 In \Fig{trajectory}(b) we have rescaled the axes of (a) by $n_c$ and $s_c$ and expanded the region near the critical point.
The detuning parameter $\Delta_s$ is the intercept of $45^{\circ}$ lines which label the trajectories of the system.
Finally, in \Fig{trajectory}(c)  (which is discussed more completely in \Sect{sec:timescale}) we have rescaled the $\Delta n/n_c$ and $\Delta s/s_c$ axes of  \Fig{trajectory}(b) by $\Delta_s$ and $\Delta_s^{(1-\alpha)/\beta}$ respectively.  
Only in (c) does the fact that the system misses the critical point by an amount $\Delta_s$ become important.

\subsubsection{Computational outline}

The goal of  the current paper is to determine how the distribution of hydrodynamic fluctuations evolves in time
as the mean entropy and baryon number densities evolve according to  \Eq{eq:dsoversvstime}, and
the system passes close to the critical point with parameter $\Delta_s$.
For several important (and related) reasons the primary object of
study is the distribution of fluctuations in the entropy per baryon $\delta \shat\equiv n \delta(s/n)$ 
\st
\label{Nss-def1}
\Nn^{\shat\shat}(t,\k) \equiv \int d^3x \, e^{i \k \cdot  (\xx - \yy)} \llangle \delta\shat(t,{\xx}) \delta\shat(t,{\yy}) \rrangle \, .
\stp
First, this correlation function 
diverges near the
critical point  as the Ising magnetic susceptibility $\chi_{\is}$, which has the largest critical exponent
$\gamma\simeq 1.23$ (see \cite{onuki2002phase} and \Sect{sec:mappingqcdtoising}). 
Second, $N^{\shat\shat}$ determines
the specific heat at constant pressure $C_p$ in the limit $\k\rightarrow 0$  (see \cite{Lifshitz:v5} and \Sect{QCDbasics}).
Finally, the $\delta \shat$ fluctuation
is a diffusive eigen-mode of the linearized hydrodynamic equations, and therefore
evolves independently of other hydrodynamic fluctuations.
The associated heat diffusion coefficient $D_{\shat}$, which controls the relaxation of $\delta \shat$,   is 
similar in magnitude to the baryon  number diffusion coefficient $D_{B}$ (see \cite{KadanoffMartin,landau1980statistical} and \Sect{sec:hydro-kinetic}).  We will determine how the amplitude and the shape of the $N^{\shat\shat}$  distribution
depend on the  parameters $\lambda$ and $\Delta_s$.

We first need to describe how this correlation function would evolve in  perfect 
equilibrium; this involves  
several ingredients as described in \Sect{sec:statics}. The time evolution of the overall
amplitude of $N^{\shat\shat}$ in equilibrium is given by $C_p(t)$ which is related  through
universality to the Ising magnetic susceptibility $\chi_\is$. In \Sect{sec:map} we
describe how to map the QCD quantities $\Delta s$ and $\Delta n$ onto the phase
diagram
of the Ising model.  Since the time dependence of $\Delta s$ and $\Delta n$ has already been prescribed in \Eq{eq:dsoversvstime}, once the QCD-to-Ising map is given, the time evolution of $\chi(t) \propto C_p(t)$ is fixed. The shape of the
$\Nn^{\shat\shat}$ distribution is controlled by the correlation length $\xi(t)$ which is also specified through universality.
In equilibrium, the relaxation time parameter $\lambda$ plays no role, and the evolution of
$\Nn_0^{\shat\shat}$ is determined only by $\tau_Q$ and  $\Delta_s$. As we show in \Sect{sec:timescale}, the relevant 
timescale for the non-trivial evolution of $C_p(t)$ and $\xi(t)$ is set by a crossing timescale:
\st
t_{\cross}  \sim  \tau_Q \Delta_s^b, \quad
b\equiv \frac{1-\alpha}{\beta}\simeq 2.7\,. 
\stp
The equilibrium evolution of the  $N^{\shat \shat}$ is summarized in
\Sect{sec:equilibrium-summary}, where the time dependence of the amplitude
$C_p(t) \propto \chi_\is(t)$ and correlation length $\xi(t)$ are shown in
\Fig{fig:chi-xi-scaling}(a) and (b) respectively. 

After specifying how the equilibrium expectation evolves we will write down a dynamical evolution equation for
$N^{\shat\shat}$ by analyzing stochastic hydrodynamics in the expanding
critical background --  see \Sect{sec:dynamics}. The diffusion coefficient
entering in this evolution equation determines a relaxation rate $\Gamma_\shat$
for the $\delta \shat$ mode, which approaches zero near the critical point,
$\Gamma_\shat \propto \xi^{-z}$  with $z = 4 -\eta$. Comparing the relaxation
rate to the rate of change of the equilibrium expectation yields an emergent
Kibble-Zurek timescale  
\st
t_{\rm kz}  \sim \tau_Q \lambda^{-a\nu z/(1 + a\nu z ) } \, ,
\quad a\equiv \frac{1}{1-\alpha} \simeq a \simeq 1.12\, ,
\stp
which  sets the 
timescale for the non-equilibrium evolution of the fluctuations.
The  Kibble-Zurek time is described more completely in \Sect{sec:KibbleZurek}. 

Our final numerical result for the time evolution of $N^{\shat \shat}$ when
the system passes directly through the critical point ($t_{\cross}=0$) is shown in \Fig{fig:tcross0} of \Sect{sec:evaluation}.
When the system misses the critical $N^{\shat\shat}(t,\k)$  generally depends on the ratio of $t_{\cross}$ and $t_{\KZ}$ 
leading to \Fig{fig:tcross1}.
Numerical estimates for the magnitude  of $N^{\shat\shat}$ and the correlation length are discussed in the conclusions.

\section{Transits of the critical point: equilibrium} 
\label{sec:statics}
In this section we will analyze the equilibrium fluctuations of $\shat$ close to critical point during a transit of
the QCD critical point. Subsequently in \Sect{sec:dynamics}  we will analyze the dynamics
of the system to determine the corresponding non-equilibrium distribution $N^{\shat\shat}$.

\subsection{Mapping the QCD equation of state onto the Ising model }
\label{sec:map}

To map the QCD equation of state  onto the Ising model, we need to relate the
temperature and chemical potential in QCD to the temperature and magnetic field
of Ising system. Alternatively we may work with extensive variables and map the 
energy and number densities of QCD to the energy density and magnetization
of the Ising model. Since the time dependence of the QCD extensive variables have already
been specified in \Eqs{eq:dsoversvstime} and \eq{eq:devstime},  the system's trajectory in the Ising phase diagram
is completely determined once this map is given.

The extensive thermodynamic variables in QCD phase diagram are denoted generically with $x^a$
\st
\label{xa-def}
  x^{a} \equiv
  \begin{pmatrix}
       e  &   n
  \end{pmatrix}\, , 
\stp
while the corresponding thermodynamically conjugate variables are denoted with capital letters
$X_{a} = -\partial s/\partial x^a$
\st
\label{eq:dsdx}
  X_{a}  = 
  \begin{pmatrix}
     -\beta  &   \muh
  \end{pmatrix} \, .
\stp
Here $\muh = \mu/T$ and  $\beta = 1/T$. Near the critical point
the entropy  can be written as a regular piece plus a singular piece\footnote{Strictly speaking it is the free energy and not the entropy which may be clearly divided into regular and singular pieces. $s_{\rm reg}$ and $s_{\rm sing}$ are
   determined from the corresponding free energies with the relation
$s=\beta p + \beta e -\muh n$, where $e$ and $n$ are derivatives of the free energy with respect to $-\beta$ and $\muh$.}, $s = s_{\rm reg}  + s_{\rm sing}$,
where  the regular piece is
\st
 s_{\rm reg} = s_c  +\beta_{c} \Delta e - \muh_c \Delta n  \, .
\stp
Then from Eq.~\eqref{eq:dsdx} the singular part of the entropy density satisfies
\st
  \Delta X_a  = -\frac{\partial s_{\rm sing} }{\partial x^a} = 
  \begin{pmatrix}
     -\Delta \beta  &   \Delta \muh
  \end{pmatrix} \, ,
\stp
with  $\Delta \beta \equiv \beta - \beta_c$ etc, so that 
\st
\label{eq:ssingqcd}
  ds_{\rm sing}(x) = -\Delta X_a(x) \, dx^a \, .
\stp

Equilibrium fluctuations in QCD are treated as in Ref.~\cite{Lifshitz:v5}.
In each subsystem  of volume $V$ which is large compared to the cube of the correlation length, the probability  of a fluctuation 
$x^a \rightarrow x^a + \delta x^a$
is Gaussian and given by
\st
P \propto  e^{\Delta S_{(2)}}  \qquad  \Delta S_{(2)} = -\frac{1}{2} V \, \S_{ab}(x) \, \delta x^a \,\delta x^b \, ,
\stp
Here the matrix $\S_{ab}$ is given  by equilibrium thermodynamics
\st
\label{Sab-def}
  \S_{ab}(x) =  \frac{\partial X_a(x)}{\partial x^b} = - \frac{\partial^2 s(x)}{\partial x^a \partial x^b } \, .
\stp
Finally if the  $\delta x(\r)$ is a function of space,  the probability becomes a functional and takes the form
\st
\label{eq:DS_2}
 P[\delta x] \propto e^{\Delta S_{(2)}}\,, \qquad \Delta S_{(2)} = -\frac{1}{2} \int d^3\r \, S_{ab}(x)  \, \delta x^a (\r)  \, \delta x^b(\r)  \, .
\stp

The extensive variables in the Ising model (the energy density and the magnetization) are denoted generically with $x^A$, distinguished from QCD case by the uppercase index:
\st
x^{A}\equiv 
\begin{pmatrix}
   \eis & \psi 
\end{pmatrix}
\, .
\stp
Here $\eis \equiv (\E -\E_c)/\tcis$ is the deviation of Ising energy density from the critical one relative to the Ising critical temperature $\tcis$, while $\psi$ is the spin density (the order parameter). 
The thermodynamically conjugate variables are $X_A=-\partial s_{\rm sing}/\partial x^A$
\st
  X_{A} =  
  \begin{pmatrix}
     \ris &   h 
  \end{pmatrix}
  \,,
\stp
where ${\ris}=(T-\tcis)/\tcis$ denotes the reduced temperature, and $h=H/\tcis$
is the reduced magnetic field (see \app{appendix:eos}).  The singular part of the Ising entropy
is
\st
\label{eq:ssingis}
 d s_{\rm sing,is}(x) = - X_A(x) dx^{A} \, ,
\stp
and  the equilibrium quadratic functional  reads
\st
   \Delta S_{(2)} = - \frac{1}{2} \int d^3\r \, \S_{AB}(x) \, \delta x^A(\r) \, \delta x^{B}(\r) \, . 
\stp


The mapping between $x^{a}$ and $x^{A}$ or $X_{a}$ and $X_{A}$ is not universal but is analytic~\cite{onuki2002phase}.
Therefore, in the vicinity of the critical point, $\Delta X_{a}$ and $X_{A}$
are related through a linear transformation specified by
2-by-2 matrix $\minv$:
\begin{eqnarray}
   \label{eq:minv}
   X_A & =& \Delta X_b \minv^{b}_{\sp A}\, ,  \qquad  \minv^b_{\sp A} = \frac{\partial X_A}{\partial X_b}\, . 
\end{eqnarray}
Similarly, the extensive variable are related with a 2-by-2 matrix $M$
\st
   x^A = M^{A}_{\sp b} \Delta x^{b}\, , 
   \qquad
    M^A_{\sp b} = \frac{\partial x^A}{\partial x^b} \, .
\stp

The matrices $M$ and $\bar M$ are inverses of each other.
Indeed, the probability of a fluctuation in the extensive QCD parameters $\delta e, \delta n$ must be the same as a corresponding fluctuation in $\delta \eis, \delta{\psi}$ in the Ising system in order to have universal behavior. 
The decrease in entropy per volume $\Delta S_{(2)}$ due to a fluctuation must be the same in both systems:
\st
\label{map-relation1}
  \delta h \, \delta{\psi} + \delta{\ris} \,  \delta{\eis}
  = \delta{\muh} \, \delta{n}  - \delta{\beta}\, \delta{e}
\stp
 i.e. $\dfluct{X}_A \dfluct{x}^A = \dfluct{X}_a \dfluct{x}^a $.  
From eq.~\eqref{map-relation1} we see that $M$ and $\minv$ are inverse matrices
of each other
\st
  \label{eq:mminv}
  M^{B}_{\sp a} \minv^a_{\sp C} = \delta^{B}_{\sp C} \, .
\stp
With this relation we also see that singular parts of the entropy differential $ds_{\rm sing}$ of the QCD and Ising systems agree.

Of the four parameters in the two-by-two matrix $\minv$ (or $M$), two of the parameters are just scale factors, while the remaining two parameters determine the directions of changing $\tau$ and $h$
in the QCD $T,\mu$ plane.
The line $h=0$ is the coexistence line in the Ising system, and must correspond to the coexistence curve, $T_{\rm cx}(\mu)$, in the QCD phase diagram.
Thus, knowledge of $T_{\rm cx}(\mu)$ places a constraint on the remaining two directional parameters of $M$, which is found by setting 
$dh=0$ (i.e. constant $h$) in \Eq{eq:minv} 
\st
    \frac{T'_{\rm cx}}{1 - (\mu_c/T_c) T'_{\rm cx}(\mu)}  \left(\frac{1}{T_c} M^{\eis}_n\right) = M^{\eis}_e  \,,
\stp
or equivalently
\st
    \frac{T'_{\rm cx}}{1 - (\mu_c/T_c) T'_{\rm cx}(\mu)}  \left(\frac{1}{T_c} \minv_{h}^{\beta}\right) = -\minv_{h}^{\muh} \, .
\stp
Following previous works~\cite{Nonaka:2004pg}, we ignore the $\mu$ dependence of $T_{\rm cx}(\mu)$ and set $T'_{\rm cx}(\mu)=0$, and thus $M^\eis_e=0$ and $\minv_{h}^{\muh}=0$.
For maximum simplicity we will also take  the direction of increasing $h$ in the Ising model to correspond with $T$ direction of QCD by setting $M^\psi_n = - M^\psi_e \mu_c$.
With these choices, the map is determined by two positive dimensionless scale factors, $(T_c M_{e}^{\psi})$ and  $-M^{\eis}_n$, leading to the definition
\begin{align}
   A_{s} \equiv& (T_c M_e^{\psi}) \, , \\
   A_{n} \equiv& -M^{\eis}_n \, .
\end{align}
The intensive parameters of the Ising model and QCD are related after elementary algebra
 \begin{align}
    \ris =&  - \frac{1}{A_n} \frac{\Delta \mu}{T_c} \,, \\
    \his =&  \frac{1}{A_s} \frac{\Delta T}{T_c} \, .
 \end{align}
In terms of the extensive parameters, this means that the QCD entropy is proportional to the order parameter
\begin{align}
   \eis =&  {-}A_n \Delta n \, , \\
   \psi  =&A_s \, \Delta s \, ,
\end{align}
where we have used $T_c \Delta s =  \Delta e - \mu_c \Delta n$. 
Finally, as discussed more completely below,  the Ising energy density and
magnetization, $(\eis,\psi)$, are determined up to two normalization constants,
$(\M_0 h_0,\M_0)$.  These constants can always be adjusted
by redefining the mapping parameters,
and we will conventionally choose  
\begin{align}
   \M_0 h_0 \equiv&  n_c \, , \\
   \M_0 \equiv&  s_c \, ,
 \end{align}
 so that the scale factors  ($A_n,A_s$)
are of order unity.
Thus, our final specification for how $(\eis, \psi)$ are related to $(\Delta n, \Delta s)$ reads
\begin{subequations}
\label{ns-mapping}
\begin{align}
   \frac{\eis}{\M_0 h_0} =&  -A_n \, \frac{\Delta n}{n_c} \,, \\ 
   \frac{\psi}{\M_0}  =&  A_s\, \frac{\Delta s}{s_c} \, . 
\end{align}
\end{subequations}
Our conclusions will be largely independent of the precise form of the mapping between QCD and Ising model. What is important in what follows is that
$A_n$ and $A_s$ are positive, dimensionless, and of order unity constants. Further
\Eq{ns-mapping} together with the time dependence of $\Delta n$ and $\Delta s$ given in \Eq{eq:dsoversvstime}
 fully specify how the QCD system evolves in the Ising model 
plane as a function of time.

\subsection{The QCD specific heat $C_p$ and the speed of sound near the critical point}
\label{sec:specificheats}

Given the Ising equation of state and the corresponding states in the QCD
medium, we may compute how the QCD specific heats and the speed of sound are related to 
the Ising susceptibilities near the critical point.  As we will review,  the critical behavior of the speed of
sound and the specific heat at constant pressure, $C_p$, are independent of the
details of the mapping matrix $M^{A}_{\Asp b}$~\cite{onuki2002phase}. $C_p$
determines the fluctuations in the entropy per baryon $\shat$, and is the most rapidly divergent equilibrium susceptibility near the QCD critical point. 

\subsubsection{The Ising model susceptibilities}
\label{Isingbasics}
The Ising model susceptibilities determine the fluctuations in the extensive quantities $x^{A}$, and are given by  the matrix
\st
  \Gi^{AB}   = \left. \frac{1}{V} \frac{\partial^2 \log Z_{\rm sing} }{\partial X_A
  \partial X_B}  \right|_{X_A=0} =  V \llangle \dfluct{x}^A \dfluct{x}^B \rrangle\, . 
\stp
The conventional names for the entries of this matrix are
   \begin{align}
      \label{susceptibility}
   \Gi^{11} \equiv & C_H \,, \\
   \Gi^{22} \equiv & \chi_\is \,, \\
   {\rm det}\, \Gi^{AB} \equiv & \chi_{\is} C_M\,,      \qquad  C_{M} = \Gi^{11} - \frac{ (\Gi^{12})^2 }{\Gi^{22} } \,,
   \end{align}
where  $C_H$ is the specific heat at constant magnetic field, and $C_M$ is the specific heat at constant magnetization.
Straightforward algebra (see \app{appendix:eos} for details) yields explicit expressions for these quantities in terms of the commonly used $R,\theta$ parametrization -- see \Eq{eq:susceptibility_explicit} in \app{appendix:eos}. As seen from the appended expressions,
the Ising susceptibility $\chi_{\is}$ and specific heat $C_M$ diverge 
as
\begin{align}
   \chi_\is  \propto& R^{-\gamma}\,,   \quad\mbox{with} \quad \gamma = 1.24
   \,, \\
   C_M \propto& R^{-\alpha}\,,   \quad \mbox{with} \quad \alpha = 0.11\, .
\end{align}
where $R\rightarrow 0$ near the critical point.
From a perspective of heavy ion collisions, the critical exponent $\alpha$ 
is so small that it will probably never be observed,  and we will focus
on susceptibility  $\chi_{\is}$.  

The inverse matrix determines the corresponding fluctuations of the intensive parameters~\cite{*[{See for example sections 16, 111, and 112:} ] [{.\; Note that $S$ in the non-relativistic literature typically denotes the entropy per particle $S/N$}] Lifshitz:v5}  
\st
 \Si_{AB} \equiv \left(\Gi^{-1}\right)_{AB} = \frac{1}{\chi_\is C_M} 
 \begin{pmatrix} \chi_\is & -\Gi^{12}  \\
 -\Gi^{12} &  C_{H}   
 \end{pmatrix} = 
     V \llangle \dfluct{X}_A \dfluct{X}_B \rrangle \,, 
\stp
which follows from the definition, $X_A = -\partial S/\partial x^A$. 
We note the correlations between the extensive and intensive variables
are simple
\st
\label{eq:corr_ext-int}
  V \llangle \delta x^A \delta X_{B} \rrangle = \delta^{A}_{B} \, ,
\stp
reflecting the relation, $\Si={\Gi}^{-1}$

Finally, let us discuss the  wavenumber dependence of the Ising correlation functions. Near the critical point the correlation function of magnetization,
\st
\llangle \psi(\kk) \psi(\kk') \rrangle \equiv \Chi(k) \,  (2\pi)^3\delta^{(3)}(\kk - \kk') \, ,
\stp
takes the form
\st
\Chi(k)  =  \chiis \, \Kchi(k\xi) \, ,
\stp
where $\Kchi(\kb)$ is a static universal function 
with unit normalization\footnote{
In principle, $\Kchi$ will be different inside and outside coexistence regime~\cite{KosterlitzPRB}. 
While including such dependence is straightforward, we will neglect this refinement in the current study. 
},
$\Kchi(0)=1$. $\Kchi$ has been studied extensively~\cite{zinn2002quantum},
and for $k\gg \xi^{-1}$ takes the asymptotic form 
\begin{align}
\label{Kchi-asym}
   \chi_\is \Kchi(\kb)=& \frac{ C_{\infty}}{k^{2-\eta}}\, ,
\end{align}
where $\eta \simeq 0.036$ is the critical exponent, and  the constant $C_{\infty}$ is independent of $\xi$. 
 We will use the simple Ornstein-Zernicke form~\cite{onuki2002phase}  
\st
\label{eq:OZform}
\Kchi(k\xi) = \frac{1}{1 + (k\xi)^{2 -\eta} } \, ,
\stp
which has the correct limits for  $k \ll \xi^{-1}$,  and  $k \gg \xi^{-1}$.


\subsubsection{The QCD susceptibilities}
\label{QCDbasics}
The corresponding QCD susceptibility matrices are
\st
  \GQCD^{ab} = \frac{1}{V} \frac{\partial^2 \log Z_{\rm sing} }{\partial X_a \partial X_b}\,,    \qquad 
  \SQCD_{ab} \equiv  (\GQCD^{ab})^{-1} \, ,
\stp
which determine the QCD fluctuations $\llangle \delta x^a \delta x^b \rrangle$,
and $\llangle \delta X_a \delta X_b \rrangle$ respectively.
The matrix  $\GQCD^{ab}$ 
determines the speed of sound $c_s^2$ and the fluctuations in
the entropy per baryon as we review below. 

To write down the formulas relating the speed
of sound to $\GQCD^{ab}$, we define derivatives of the pressure
\begin{align}
   \label{pdefs}
p^{a} &\equiv \frac{\partial p}{\partial X_{a}}, \quad
   (p^e, p^n) =
   \left(-\frac{\partial p}{\partial \beta}, \frac{\partial p}{\partial \hat\mu} \right) 
   = \left( \frac{w}{\beta} , \frac{n}{\beta}\right) \, ,
\end{align}
and then the speed of sound, $c_s^2 = \left(\partial p/\partial e\right)_{n/s}$, is given by 
\begin{subequations}
\begin{align}
   c_s^2 &=\left(\frac{\partial p}{\partial e}\right)_n +\frac{n}{w}\left(\frac{\partial p}{\partial n}\right)_e\,, \\
   \label{cs2p2}
&=\frac{\beta}{w}p^{a} \SQCD_{ab} p^{b}.
\end{align}
\end{subequations}
As usual, $w\equiv e+p$ is the enthalpy density.
From this expression we see that fluctuations in the pressure $\dfluct{p} = p^a \dfluct{X}_a$ determine the speed of sound
\st
 \label{pflucts1}
 V\llangle \dfluct{p}^2 \rrangle = p^a \SQCD_{ab} p^b  = \frac{w c_s^2}{\beta}  \, .
\stp

The fluctuations in the entropy per baryon 
will play a central role in what follows, and thus 
we define
\st
 \delta \sh \equiv n \delta\left(\frac{s}{n}\right) = \delta s - \frac{s}{n} \delta n \, .
\stp
The fluctuations in $\sh$ can be written in terms of $\delta e$ and $\delta n$
\st
T \delta \sh =  \delta  e  -  \frac{w}{n} \delta n   \, ,
\stp
and are uncorrelated with the fluctuations in the pressure
\st
\label{eq:corr_pandsh}
 \llangle \delta p \, \delta \sh \rrangle = 0   \, ,
\stp
which can be derived from \Eq{eq:corr_ext-int} and \Eq{pdefs}. A more complete discussion 
of this and the thermodynamic relations  in
the rest of this section
is given in Refs.~\cite{Lifshitz:v5,onuki2002phase}.
The  fluctuations in $\sh$ are determined by 
the specific heat at constant pressure, $C_p \equiv  n T \left(\frac{\partial(s/n) }{\partial T} \right)_p$, via
\st\Nss\equiv
\label{shatflucts}
V\llangle (\delta\sh)^2  \rrangle =  V \llangle (\delta s - \frac{s}{n} \delta n)^2 \rrangle
        =C_p \, .
\stp
Straightforward analysis shows that $C_p$ is related to determinant of the susceptibility matrix
\st
\label{cp2gab}
\left(\frac{n T}{w} \right)^2 C_p =  \frac{ \beta c_s^2 }{w} \det \GQCD^{ab} \, .
\stp
The specific heat at constant pressure is also related to the specific heat at constant 
volume $C_V \equiv T (\partial s/\partial T)_n$ through 
the familiar relation 
\st
\label{CpandCv}
\left( \frac{n T}{w} \right)^2 C_p =  T\frac{\partial n}{\partial \mu} \, \left( \frac{ TC_V c_s^2 }{w} \right) \, .
\stp
In the low density limit, $n\rightarrow 0$, the final factor on the
r.h.s. approaches unity,  $\left(TC_V c_s^2/w\right){\rightarrow} 1$.
Eq.~\eqref{CpandCv}  leads to an important relation, \Eq{eq:dshatdb} below,  between the baryon
number diffusion coefficient and the diffusion coefficient of $\shat$.

In practice, both theoretically  and experimentally,  it is easier to work with  the correlation function of $\shat$ 
rather than  fluctuations  of  $\shat$ in a finite  volume $V$
\st
\label{Nss-def}
\Nn^{\shat\shat}(t,\k) \equiv \int d^3x  \, e^{i \k \cdot  (\xx - \yy)}\,  \llangle \delta\shat(t,{\xx}) \delta\shat(t,{\yy}) \rrangle \, .
\stp
In equilibrium, \Eq{shatflucts} predicts that  $\Nn^{\shat\shat}(t,\k)$ approaches $C_p$ as $\k\rightarrow 0$.




\subsubsection{QCD fluctuations near the critical point}
\label{sec:mappingqcdtoising}

We have now specified how the speed of sound and specific
heats are related to the QCD susceptibility matrix $\GQCD^{ab}$.
The QCD susceptibilities are related to the corresponding Ising quantities with the mapping matrices of \Sect{sec:map}.
\st
  \GQCD^{ab} = \minv^{a}_{\;A} \, \minv^{b}_{\;B} \, \Gi^{AB} \, .
\stp
As we will now review,
near the critical point the speed of sound, $c_s^2$, approaches zero  as 
 $C_M^{-1} \propto R^{\alpha}$, while  the specific heat, $C_p$,
diverges as $\chi_\is \propto R^{-\gamma}$~\cite{onuki2002phase}.  
This is independent of the details of the mapping matrix $M^{A}_{\Asp b}$.
From a practical perspective
 this means that the softening of the equation of state near the critical
point  will probably be too small to observe (since $\alpha$ is small), and the experimental heavy ion program
should  focus on the fluctuations in $s/n$ which reflects the diverging value
of the specific heat $C_p \propto \chi_\is$.

To review how the speed of sound behaves near the Ising critical point, we first note that  
by inserting unity of the form  $ \minv^{a}_{\;B} M^{B}_{\;c}= \delta^{a}_{c}$ 
into \Eq{pflucts1}, we can express the speed of 
sound near the critical point as 
\begin{equation}
\label{speedofsound}
c_s^2 = \frac{\beta}{w} p^A S_{AB} p^B   \simeq  
\frac{\beta}{w} \left(\frac{\partial p}{\partial r} \right)^2  
  \frac{1}{C_M} \, , 
\end{equation}
where we define $p^{A} = (\partial p/\partial X_A)$ and thus
$ \left(\partial p/\partial r\right)_{h} = M^\epsilon_{n} p^n + M^{\epsilon}_{e} p^e$, 
is derivative of the QCD pressure in the direction of reduced Ising temperature. We note
that $\left(\partial p/\partial r\right)$ remains finite near the critical point.
In approximating \Eq{speedofsound},
we recognized that near the critical point $\chi_\is$ is strongly divergent, and thus the $rr$ component in $p^A S_{AB} p^B$  dominates
the sum.
This shows (as claimed) that the speed of sound approaches zero like the Ising specific heat $C_M^{-1}$, i.e. as $R^{\alpha}$.
In the case of the simple mapping  described in \Sect{sec:map} we have
\st
c_s^2 = 
\frac{\beta}{w}  \frac{ (T_c n_c A_n)^2 }{C_M} 
\,. 
\stp
In the rest of this paper we will focus on the specific heat $C_p$ which exhibits a much more dramatic behavior, diverging as $R^{-\gamma}$ near the critical point.

The behavior of $C_p$ near the critical point is determined by the determinant  in \Eq{cp2gab} and
the relation between the determinants of the QCD and Ising systems
\st
\det \GQCD^{ab} = ({\rm det} \bar M)^2 \, {\rm det} \Gi^{AB}
\stp
Thus since $\det \Gi^{AB} = \chi_\is C_{M}$, we find with Eqs.~(\ref{cp2gab}) and (\ref{speedofsound}) that 
\st
C_p = \frac{\left(\frac{1}{T_c n_c} \frac{\partial p}{\partial r} \right)^2  }{ (T_c \,{\rm det} M)^2} \,  \chi_\is \, .
\stp
The factors $(T_c\,{\rm det} M)$ and $\left(\partial p/\partial r\right)/ n_cT_c$ are both dimensionless and of order 
unity.  Thus, independently of the details between the QCD and Ising variables, the specific heat $C_p$ is proportional to the Ising susceptibility $\chi_{\rm is}$ and diverges as $R^{-\gamma}$.
For the simple mapping of \Sect{sec:map} the specific heat takes the particularly simple form
\st
\label{cp-Chi}
C_p  =  \frac{\chi_\is}{A_s^2 } \, ,
\stp
which we will assume in what follows. 

Finally, later we will study the correlation function $\Nn^{\shat\shat}(t,\k)$ 
as a function of $\k$. In equilibrium, this will take the form 
\st
\label{cp-Chi-k}
\Neq^{\shat\shat}(t,\k)  = \frac{\chi_\is}{A_s^2 } \frac{1}{1 + (k\xi)^{2-\eta}} \,,
\stp
where we have adopted for simplicity Ornstein-Zernicke form, which has the properties discussed in \Sect{Isingbasics}.

At this point we need to determine how the parameters $\chi_\is(t)$ and $\xi(t)$ depend on time  when $\Delta n$ and $\Delta s$ follow the adiabatic trajectory
parametrized by \Eq{eq:dsoversvstime}. We will turn to this task in the next section.



\subsection{The timescale for the scaling regime during a transit of the  QCD critical point}
\label{sec:timescale}

We have now specified how the extensive Ising variables
$(\eis,\psi)$ are determined by the QCD quantities
$(\Delta n, \Delta s)$ with \Eq{ns-mapping}. We also have specified how the extensive QCD quantities depend on time in \Eq{eq:dsoversvstime}.
Finally, the Ising equation of state determines the time dependence of the
corresponding susceptibilities and correlation lengths,
from the time dependent extensive Ising variables.  In this section we will
show how the scaling form of the Ising equation of state leads to a characteristic scaling form
in time for these quantities.


Outside of the coexistence region, the scaling of the Ising equation of state implies the following scaling forms
for the extensive variables $(\eis,\psi)$ as a function of $(r,h)$
\begin{subequations}
   \label{isingscalingeos}
\begin{align}
   \eis =& \mathcal M_{0} h_0 \, |r|^{1-\alpha}f_{\eis} (z)\,, \\
   \psi  =& \mathcal M_{0} \,  |r|^{\beta}f_{\psi}({z}) \, ,
\end{align}
\end{subequations}
Here $(\M_0 h_0, \M_0)\equiv (n_c, s_c)$ are two (conventional) constants described above, 
and
below $f_X(z)$ denotes a generic universal scaling function of the variable $z\propto r/|h|^{1/\beta\delta}$ (see \Eq{zdef} in \app{appendix:eos} for a complete definition of $z$.)    
 All susceptibilities and correlation lengths take 
this generic form, and no additional constants
need to be introduced.
In practice, given $(\eis,\psi)$ we numerically determine $(R,\theta)$ from the  Ising parametrization described in \app{appendix:eos}, and then evaluate 
all other thermodynamic functions.

As $z \rightarrow z_0\equiv-\infty$, the system  approaches the 
coexistence region,
and $f_{\eis}(z) $ and $f_{\psi}(z)$ approach  $-1$ and $1$ by convention\footnote{
A handy Mathematica notebook which evaluates all universal Ising thermodynamic variables and 
correlation lengths is made available as part of this work.}.  
Inside the coexistence region 
the energy density is related to the temperature by
\begin{align}
   \eis =& \mathcal -\M_0 h_0 |r|^{1-\alpha}  \, , 
\end{align}
and the magnetization lies in the range $(-\psi_0, \psi_0)$ where
\begin{align}
   \psi_0 =& \mathcal \M_0 |r|^{\beta}   \, .
\end{align}


These expressions for the extensive quantities in terms of
the intensive ones may be inverted. We define a new scaling variable
based on extensive variables
\st
\label{u-def}
u \equiv \frac{\eis}{\M_0 h_0} \left(\frac{\M_0}{|\psi|} \right)^{b} \, ,
\stp
where $b=(1-\alpha)/\beta\simeq 2.7$, and then  outside
the coexistence region
\begin{align}
   \label{uscalingeos}
   r =& \left(\frac{|\eis|}{\mathcal M_0 h_0}\right)^{a} f_r(u) \, ,\\
   z   =& f_{z}(u) \,,
\end{align}
with $a=1/(1-\alpha)\simeq 1.12$.  The 
system is in the coexistence region for  $u  < -1$.

The advantage of a scaling variable based on extensive quantities is that the extensive quantities depend on time in a simple way. 
Indeed, the scaling variable $u$ is approximately linear in time
\st
\label{usimple}
u =   \frac{A_n \, t/\tau_Q}{(A_s\, (\Delta_s - t/\tau_Q))^b }  \simeq \frac{A_n}{A_s^b} \,  \frac{t}{\tau_Q\Delta_s^b} \, .
\stp
In the last step, we recognized that in order to see the detailed scaling
structure in the  equation of state (which is parametrized by $f_{r}(u)$ in
\eqref{uscalingeos}), we must have $|u| \sim |z|  \sim  |\theta| \sim 1$.  For $|u| \sim 1$,
$|t/\tau_Q| \sim \Delta_s^b$ and is  small compared $\Delta_s$ in this regime
this regime.
From the last equality of \Eq{usimple}, the system crosses the detailed scaling regime  over a time period  of order 
\st
  t_{\cross} \sim \tau_Q \Delta_s^b \, .
  \stp
 Parametrically outside of this time window the scaling functions such as $f_{r}(u)$ may  be treated as constants.
 Inside of this time window the QCD parameters are of order
 \st
 \frac{\Delta s}{s_c} \sim \Delta_s\,,  \qquad \frac{\Delta n}{n_c} \sim  \Delta_s^b \,.
 \stp
 Accordingly, in \Fig{trajectory}(c) we have rescaled the $x$ and $y$ axis by $\Delta_s^b$ and $\Delta_s$, which flattens the $45^{\rm o}$ trajectory lines in 
 \Fig{trajectory}(b).   It is only in this regime that
 the detailed scaling structure of the Ising equation of state (as recorded by the $(R,\theta)$ parametrization) is really necessary.

 To simplify notation  we absorb the mapping constants 
 into the definition of the parameters defining
 \begin{align}
    \overline{\tau}_{Q} \equiv& \frac{\tau_Q}{A_n}\,, \\
    \overline{\Delta}_s \equiv&  A_s \Delta_s\,.
\end{align}
The crossing time is  defined  as the time when the system
leaves the coexistence region (see \Fig{trajectory}(c))
\st
   \label{eq:tcross}
   t_{\cross} \equiv  -\bar \tau_Q\,  \bar  \Delta_s^b\,,  \qquad \mbox{with} \qquad u = \frac{t}{|t_{\cross}|}\,,
   \stp
   so $u=-1$ corresponds to $t=t_{\crs}$.  The Ising energy and
   order parameter have a simple time dependence
   \st
      \epsilon = \M_0 h_0 \, \frac{t}{\bar \tau_Q} \, ,
      \qquad \psi =  \M_0 \, \bar \Delta_s \, .
   \stp

\begin{figure}
   \begin{center}
   \includegraphics[width=0.49\textwidth]{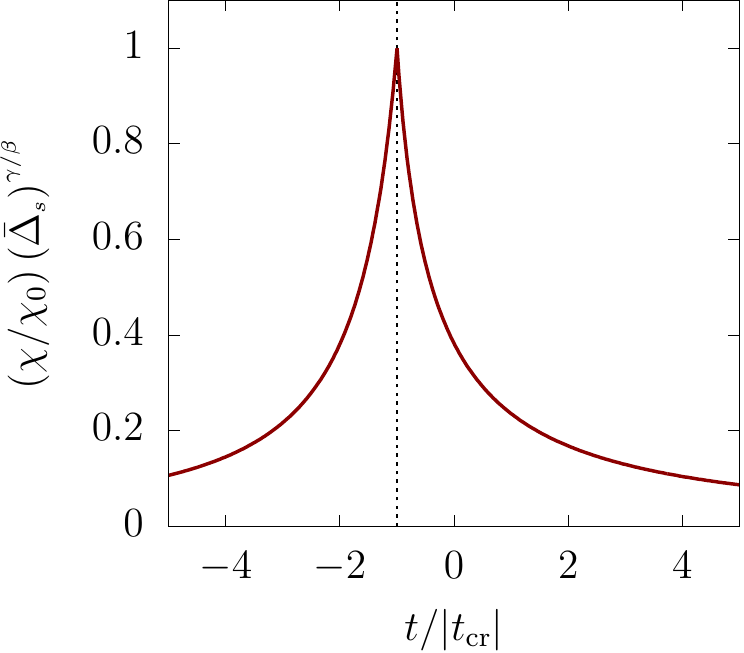}
   \includegraphics[width=0.49\textwidth]{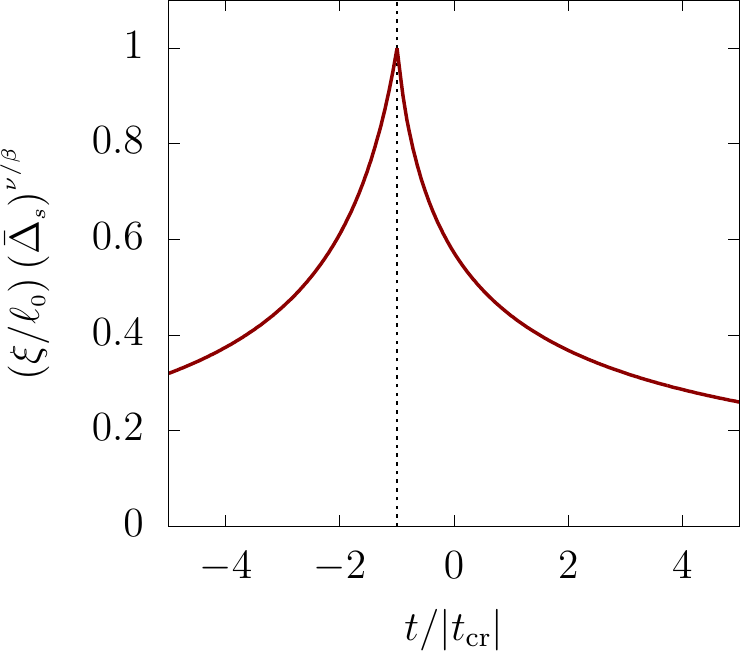}
   \end{center}
   \caption{
\label{fig:chi-xi-scaling}
The Ising susceptibility and correlation length 
as a function of time during a transit
of the QCD critical point along an adiabatic trajectory characterized by $\bar \Delta_s$. The time axis has been rescaled by $t_{\cross} \sim \tau_Q \Delta_s^b$, see \Eq{eq:tcross}.
The $y$ axes have been rescaled by an appropriate power of $\bar{\Delta}_s$ 
so that the curve is independent of $\bar \Delta_s$.
   }
\end{figure}

The scaling of the Ising susceptibility and other thermodynamic
quantities with with $\eis$ and $u$ imply a specific scaling in time. 
For instance, using the Ising parametrization in \app{appendix:eos}, the susceptibility behaves as
\st
\label{chitdependence}
\chi_\is = \chi_0  \left( \frac{|\eis|}{\M_0 h_0} \right)^{-a\gamma}  f_{\chi} \left(u\right) \, ,
\stp
where 
\st
\chi_0  \equiv 0.365\frac{s^2_c}{n_c} \, .
\stp
is the typical size of $C_p$ away from the critical point, and
we recall that $s^2_c/n_c = \M_0/h_0$.
The scaling function is continuous and takes the form
\st
f_\chi\left(u\right) =
\begin{cases} 
   1    & u  < -1 \\
   f_{\chi}\left(u\right)    & u > -1
\end{cases} \,, 
\stp
with limiting values
\begin{gather}
   f_{\chi}(-1) = 1\,,  \qquad f_{\chi}\left(u\right)  \xrightarrow[{u\rightarrow\infty}]{} f_{\chi}^+ \equiv 1.954 \, . 
\end{gather}
The combination $|u|^{-a \gamma} f_{\chi}\left(u\right)$ 
is regular and decreasing for $u> -1$. 
Thus, the equilibrium susceptibility as a function of time  takes the following  form
\begin{subequations}
\begin{align}
   \chi_\is =& \chi_0  \,  \left(\frac{|t|}{\overline{\tau}_Q }\right)^{-a\gamma} f_{\chi}\left(\frac{t}{|t_{\crs}|} \right)\, ,  
\end{align}
which can be written as function $t/|t_{\crs}|$ using \Eq{eq:tcross} 
\begin{align}
   \label{eq:chiis2}
   \chi_\is=& \chi_0 \left(\overline{\Delta}_s\right)^{-\gamma/\beta}   \, \left|\frac{t}{t_{\crs}} \right|^{-a\gamma}  f_{\chi}\left(\frac{t}{|t_{\crs}|} \right) \, .
\end{align}
\end{subequations}
Eq.~\ref{eq:chiis2}  is plotted in \Fig{fig:chi-xi-scaling}(a).
To evaluate $|u|^{-a\gamma} f_{\chi}(u)$ in practice,
we determine the $(R,\theta)$ associated with $(\epsilon, u)$ numerically
-- see \app{appendix:eos}.

The correlation length follows a similar pattern. The equilibrium correlation length in the Ising model takes the scaling form (see \app{appendix:eos})
\st
\label{xitdependence}
\xi(t) = \ell_o \left( \frac{|\eis|}{\M_0 h_0} \right)^{-a\nu}  f_{\xi} \left(u\right) \, ,
\stp
where 
\st
\ell_0 \equiv 0.365  \, n_c^{-1/3} \,,
\stp
is of order the inter-particle spacing, and we recall that $\M_0 h_0 = n_c$.
The  limiting values of the analogous scaling function $f_{\xi}(u)$ are 
\begin{subequations}
   \begin{gather}
f_{\xi}(-1) = 1 \, , 
\qquad
f_{\xi}\left(u\right)  \xrightarrow[{u\rightarrow\infty}]{} f_\xi^+ \equiv 1.222 \, ,
\end{gather}
\end{subequations}
and
$|u|^{-a \nu} f_{\xi}\left(u\right)$ is regular and decreasing for $u> -1$.
The equilibrium correlation length as a function of time  takes  form
\begin{subequations}
\begin{align}
   \xi(t) =& \ell_0  \,  \left(\frac{|t|}{\overline{\tau}_Q
   }\right)^{-a\nu} f_{\xi}\left(\frac{t}{|t_{cr}|} \right)\, ,  
\end{align}
or after using the definition of $t_{\rm cr}$ (\Eq{eq:tcross})
\begin{align}
   \label{eq:xiis2}
   \xi(t)=&   \ell_0 \left(\overline{\Delta}_s\right)^{-\nu/\beta}   \, \left|\frac{t}{t_{\crs}} \right|^{-a\nu}  f_{\xi}\left(\frac{t}{|t_{\crs}|} \right) \, .
\end{align}
\end{subequations}
Eq.~\ref{eq:xiis2}  is plotted in  \Fig{fig:chi-xi-scaling}(b).
To evaluate $|u|^{-a\nu} f_{\xi}(u)$ in practice 
we use the numerical data on the Ising model from  Engels, Fromme and Seniuch~\cite{Engels:2002fi} -- see \app{appendix:eos}.

\subsection{Summary of the equilibrium expectation}
\label{sec:equilibrium-summary}

To conclude this section let us collect and review the equilibrium formulas. $N^{\shat\shat}(t,\k)$ in equilibrium
takes the approximate form, from Eqs.~\eqref{cp-Chi} and \eqref{cp-Chi-k},
\st
\label{eq:equilibrium-summary}
\Nn_{0}^{\shat\shat}(t, \k) = \frac{1}{A_s^2} \frac{\chi_\is(t) }{1 + (k\xi(t))^{2-\eta} } \, .
\stp
where $A_s$ is a constant determined by the mapping between QCD and the Ising model.
The specific heat and equilibrium correlation length are universal functions of time as shown in \Fig{fig:chi-xi-scaling}, 
and the timescale for their evolution is set by $t_{\cross} \sim \tau_Q \Delta_s^b$.
In the next section we will describe how the system evolves according to stochastic hydrodynamics, and tries  to approach this time dependent equilibrium expectation.

\section{Transits of the critical point: dynamics}
\label{sec:dynamics}

The primary purpose of this work is to discuss the fluctuations of thermodynamic variables (e.g. $e,n$) for a system transiting close to the QCD critical point. 
Specifically, we will focus on the time evolution of the correlation functions of the thermodynamic variables, which quantify the fluctuations with a specific wave number $k$.
In the previous section, 
we have analyzed the equilibrium behavior of these correlations,
and now we will study their dynamical evolution.

We first determine  this evolution in the hydrodynamic regime, $k \ll \xi^{-1}$.  
To this end, we start from fluctuating hydrodynamics, and
 derive a set of relaxation  equations 
for the correlations, which we refer to as the hydro-kinetic equations~\cite{Akamatsu:2016llw}.
In the previous section, 
we showed that 
critical fluctuations are more enhanced in the $\shat$ mode than in  any other
combination of thermodynamic variables. 
When we apply the hydro-kinetic equations (\Eq{eq:hydrokinetic_equil} below) to a system near a critical point, 
we find that the equilibration of the $\shat$ correlator $N^{\shat\shat}$ is independent of the other hydrodynamic modes, 
allowing us to focus on $N^{\shat\shat}$.
%
%

The description of $N^{\shat\shat}$ near a critical point, even in equilibrium, involves an additional length scale. 
As we have seen in \Eq{eq:equilibrium-summary}, 
the behavior of $N^{\shat\shat}$ in equilibrium exhibits a non-trivial dependence on the wavenumber $k$, and such dependence is characterized by the correlation length $\xi$. 
To model the off-equilibrium evolution of $N^{\shat\shat}$ in the  scaling region,
we need to extend the hydro-kinetic equations
to larger $k$, $\xi^{-1} \lesssim  k \ll \ell_{0}^{-1}$.  This
is done schematically in \Sect{sec:kxi1} -- see \Eq{master}.  
It should be made clear that 
\Eq{master} is simply a rough model we will use to describe
the dynamics of $N^{\shat\shat}$ in the scaling 
regime, and we defer  a systematic treatment to future work.
In \sec~\ref{sec:KibbleZurek}, we estimate the characteristic time and length scales of $N^{\shat\shat}$. 
Finally, we evaluate $N^{\shat\shat}$ numerically by solving \Eq{master} numerically
to determine the time evolution fluctuations during a transit of the critical point. 


\subsection{The evolution of fluctuations for a fluid with finite baryon density}

\subsubsection{The derivation of hydro-kinetic equations}
\label{sec:hydro-kinetic}
We begin by considering the fluctuations around a uniform static fluid background of the extensive thermodynamic variables
$e(t,{\bm x}) = e + \delta e(t,\bm x)$, $n(t,{\bm x})=n + \delta n(t,{\bm x})$,
and momentum $\vec g(t,{\bm x}) \equiv w\vec u(t,{\bm
x})$, where $\vec{u}(t,\vx)$ denotes the fluid velocity.
In  $k$-space,
the fluctuations of longitudinal momentum $g \equiv \vec g \cdot\hat{k}$ will mix with $\delta e, \delta n$ at finite density, and we will denote them collectively as
\footnote{The bar in $x^{\bar a}$ and $X^{\bar a}$
indicate that the longitudinal momentum and velocity are appended to the set
$x^a$ and $X^{a}$ defined in \Sect{sec:map}}:
\begin{align}
\d x^{\bar a} \equiv \(\d e, \d n, g\)\, . 
\end{align}
Transverse components of the momentum, $\vec g_T\cdot \vec k = 0$, decouple from $\d x^{\bar a}$ modes in the linear regime (see \Eq{hydro-lin} below). 

We are interested in the equal-time correlation function $\Nn^{\bar a\bar b}(t,{\bm k})$ in $k$-space:
\begin{align}
\label{Nab-def}
\langle \delta x^{\bar a}(t,{\bm k})\delta x^{\bar b}(t,-{\bm k'})\rangle &\equiv (2\pi)^3\delta^{(3)}({\bm k}-{\bm k'})\Nn^{\bar a\bar b}(t,{\bm k})\, ,
\end{align}
The equilibrium values of $N^{\bar a\bar b}$, namely $N^{\bar a\bar b}_{0}$, 
are given by the  susceptibility matrix:
\begin{eqnarray}
N^{\bar a\bar b}_{0}= \(\S_{\bar a\bar b}\)^{-1}\, , 
\end{eqnarray}
where
\begin{eqnarray}
\mathcal S_{\bar a\bar b} &= \begin{pmatrix}
\mathcal S_{ee} & \mathcal S_{en} & 0 \\
\mathcal S_{ne} & \mathcal S_{nn} & 0 \\
0 & 0 & \frac{\beta}{w}
\end{pmatrix}\, ,
\end{eqnarray}
and where $\S_{ee}, \S_{en}, \S_{nn}$ are defined in \Eq{Sab-def}.

In order to derive a relaxation equation for $\Nn^{\bar a\bar b}(t,{\bm k})$, we consider the linearized stochastic hydrodynamic equations in the $k$-space:
\begin{subequations}
\label{hydro-lin}
\begin{align}
 \frac{\partial}{\partial t} \delta e(t,{\bm k}) &=- i\vec  k\cdot \vec g\, ,\\
 \frac{\partial}{\partial t}\delta n (t,{\bm k})&= -\frac{n}{w}i\vec k\cdot\vec g
- \lambda_{B}\, T k^2 \delta\hat\mu -\xi_n\, ,\\
 \frac{\partial}{\partial t} \vec g(t,{\bm k})  &= -i\vec k \delta p - \frac{\eta k^2}{w} \vec g
-\frac{\zeta + \frac{1}{3}\eta}{w}\vec k (\vec k\cdot \vec g) -\vec \xi\ .
\end{align}
\end{subequations}
The noise terms are introduced above to describe dynamics of hydrodynamic fluctuations\footnote{We use the Landau fluid frame throughout, and therefore the noise is absent in the first equation of \Eq{hydro-lin}. 
}, 
and the noise correlations are constrained by the fluctuation-dissipation theorem (see for example Ref.~\cite{Lifshitz:v5}):
\begin{subequations}
\begin{align}
\langle\xi^i(t,{\bm k})\xi^j(t',-{\bm k'})\rangle
&=2T\left[
\eta\, k^2\delta^{ij} + \left(\zeta+\frac{1}{3}\eta\right)k^ik^j
\right](2\pi)^3\delta^{(3)}({\bm k - \bm k'})\delta(t-t')\, ,\\
\langle\xi_n(t,{\bm k})\xi_n(t', -{\bm k'})\rangle
&=2T \lambda_{B}\, k^2 (2\pi)^3\delta^{(3)}({\bm k - \bm k'})\delta(t-t')\, ,\\
\langle\xi^i(t,{\bm k})\xi_n(t',-{\bm k'})\rangle &= 0\, .
\end{align}
\end{subequations}
As usual, shear viscosity, bulk viscosity and baryon conductivity are denoted by $\eta,\zeta,\lambda_{B}$ respectively.

From the hydrodynamic equation \eqref{hydro-lin}, we write the equation for $x^{\bar a}$ in a compact fashion:
\begin{subequations}
\label{hydro-lin-x}
\begin{align}
\frac{\partial}{\partial t} \delta x^{\bar a}(t,{\bm k}) =& -ik \mathcal L^{\bar a\bar b} \delta X_{\bar b} + k^2\Lambda^{\bar a\bar b} \delta X_{\bar b} + \xi^{\bar a}\, , 
\nonumber \\
=&
 -ik L^{\bar a}_{\,\,\,\, \bar b} \delta x^{\bar b} + k^{2}\, {\cal D}^{\bar a}_{\,\,\,\,\bar b}\, \delta x^{\bar b} + \xi^{\bar a}\, ,
 \end{align}
with noise correlator
\begin{align}
\langle\xi^{\bar a}(t, {\bm k})\xi^{\bar b}(t', -{\bm k'})\rangle =& 2k^2\Lambda^{\bar a\bar b}(2\pi)^3 \delta^{(3)}(\vk-\vk')\delta (t-t')\, .
\end{align}
\end{subequations}
Here the matrices are
\begin{align}
\label{calL-Lambda}
\mathcal L^{\bar a\bar b} = \begin{pmatrix}
0 & 0 & p^{e} \\
0 & 0 & p^{n} \\
p^{e} & p^{n} & 0
\end{pmatrix},\quad
\Lambda^{\bar a\bar b} = T 
\begin{pmatrix}
0 & 0 & 0 \\
0 & \lambda_{B} & 0 \\
0 & 0 & \zeta+\frac{4}{3}\eta
\end{pmatrix}\, ,
\end{align}
with $\(p^{e},p^{n}\)=\(w/\b, n/\b\)$  defined in \Eq{pdefs}.
Generalizing the discussion in \Sect{sec:map}, 
we have introduced conjugate variables through the relation $\delta X_{\bar a}  = \mathcal S_{\bar a\bar b} \delta x^{\bar b}$ 
\begin{align}
X_{\bar a}\equiv \(-\beta, \hat \mu, \frac{\beta g}{w} \)\, .
\end{align}
In the second line of \Eq{hydro-lin-x}, we have further defined:
\begin{eqnarray}
L^{\bar a}_{\,\,\,\, \bar c}&\equiv & \mathcal L^{\bar a\bar b}\, {\cal S}_{\bar b \bar c}\, , 
\\
\label{D-S}
{\cal D}^{\bar a}_{\,\,\,\, \bar c}&\equiv & \Lambda^{\bar a\bar b}\, {\cal S}_{\bar b \bar c}\, .
\end{eqnarray}

By carefully averaging out the noise, 
we obtain the following equation for $N^{\bar a\bar b}$ from \Eq{hydro-lin-x}
\begin{align}
\label{hydro-kinetic}
\frac{\partial}{\partial t} \Nn(t,{\bm k}) =&
 -ik(L \cdot \Nn - \Nn\cdot L^{T})
-k^2({\cal D}\cdot \Nn + \Nn\cdot\mathcal {\cal D}^{T})+ 2k^2\Lambda\, 
\nonumber\\
 =&-ik(L \cdot \Nn - \Nn\cdot L^{T})
-k^2({\cal D}\cdot \Nn + \Nn\cdot {\cal D}^{T})
+ k^{2}\,\({\cal D}\cdot \Nn_{0} + \Nn_{0}\cdot {\cal D}^{T}\),
\end{align}
where in the second line of \Eq{hydro-kinetic}, we have used the relation~\eqref{D-S} and $\Neq=\mathcal S^{-1}$. 
The last term on the R.H.S. of \Eq{hydro-kinetic} arises from the noise $\xi^{\bar a}$ and acts as a source. 
The correlations will propagate and dissipate, as described by the first and second terms on the R.H.S. of \Eq{hydro-kinetic} respectively. 
When $N=\Neq$, 
the propagation term vanishes, i.e.
$L\cdot N_{0}-N_{0}\cdot L=0$, 
and the last two terms on the R.H.S of \Eq{hydro-kinetic}  balance with each other. 
Therefore, $\Neq$ is a static solution to \Eq{hydro-kinetic} as it should be.


Following Ref.~\cite{Akamatsu:2016llw} and for later convenience, 
we will consider the fluctuations in $\delta x^{(\alpha)}$,
which is given by a specific linear combination of $\delta x^{\bar a}$, 
namely $\delta x^{(\alpha)} \equiv \delta x^{\bar a} e^{(\alpha)}_{\bar a}$. 
Here
$ e^{(\alpha)}_{\bar a}$ is defined as the left eigenvectors for the non-hermitian matrix $L$:
\begin{align}
\sum_{\bar a}\, e^{(\alpha)}_{\bar a}\, L^{\bar a}_{\,\,\,\, \bar b} = \lambda^{\alpha}e^{(\alpha)}_{\bar b}\, ,\qquad
\sum_{\bar b}\, L^{\bar a}_{\,\,\,\, \bar b}\,  e^{\bar b}_{(\alpha)} = \lambda^{\alpha}\,e^{\bar a}_{(\alpha)}\, ,
\end{align}
where $\l^{\a}$ are corresponding eigenvalues, and where we have also introduced right eigenvectors $e^{\bar a}_{(\alpha)}$. 
Here $e^{\bar a}_{(\alpha)}$ and $e^{(\alpha)}_{\bar a}$ satisfy the  orthogonality relations:
\begin{align}
\sum_{\bar a}\, e^{(\alpha)}_{\bar a}\,e^{\bar a}_{(\beta)} &= \delta^{\alpha}_{\, \beta}\, , \quad
\sum_{\alpha} e^{\bar a}_{(\alpha)}e_{\bar b}^{(\alpha)} =  \delta^{\bar a}_{\,\bar b}\, .
\end{align}
Consequently, $L^{\a}_{\,\,\,\,\b}$ is diagonalized as 
\begin{eqnarray}
\label{L-dia}
L^{\alpha}_{\,\,\,\,\beta} \equiv e^{(\alpha)}_{\bar a}  L^{\bar a}_{\,\,\,\,\bar b} e^{\bar b}_{(\beta)}
=\l^{\a}\, \delta^{\a}_{\b}\, . 
\end{eqnarray}
We denote the three eigen-modes by $\alpha = +, -, \shat$ for reasons which will become obvious shortly. 
In what follows, 
we will consider the correlation functions of those modes:
\begin{align}
\label{Nab-def2}
\langle \delta x^{\a}(t,{\bm k})\delta x^{\b}(t,-{\bm k'})\rangle &\equiv (2\pi)^3\delta^{(3)}({\bm k}-{\bm k'})\Nn^{\a\b}(t,{\bm k})\, . 
\end{align}

%
%

To better understand the physical meaning of $\d x^{(\a)}$, we write down the eigenvalues 
\begin{eqnarray}
\label{eigenvalue}
%
\l^{\pm} =\pm c_{s}\, , 
\qquad
\l^{\shat} =0\, ,
\end{eqnarray}
and specific form of the eigenvectors:
\begin{subequations}
\label{basis}
\begin{align}
e_{(\pm)} &= 
\frac{1}{\sqrt{2}}
\begin{pmatrix}
1 \\ \frac{n}{w} \\ \pm c_s
\end{pmatrix}\, ,
\quad
e_{(\shat)} = 
 \frac{nT}{c_s^2w}
 \begin{pmatrix}
\frac{\partial p}{\partial n} \\ -\frac{\partial p}{\partial e} \\ 0
\end{pmatrix},\\
e^{(\pm)} &= 
\frac{1}{\sqrt{2}c_s^2}
\left(
\frac{\partial p}{\partial e}, \frac{\partial p}{\partial n}, \pm c_s
\right)\, , 
\quad
e^{(\shat)}= 
\left(
   \frac{1}{T}, -\frac{w}{nT}, 0
\right).
\end{align}
\end{subequations}
Consequently, 
\begin{eqnarray}
\delta x^{(\pm)} =\frac{1}{\sqrt{2}c_s^2}(\delta p \pm c_s g)\, , \qquad
\delta x^{(\shat)} = \delta \shat\, .
\end{eqnarray}
It should be clear now that those two modes with eigenvalues $\pm c_{s}$
correspond to two propagating sound modes, and the mode with
zero eigenvalue is identical to the $\shat$ mode. 
To find the equilibrium variances of these fluctuations 
we evaluate $N^{\a\b}_{0}=e^{(\a)}_{\bar a}N^{\bar a\bar b}_{0}\,
e^{(\b)}_{b}$ and find the non-zero components
\begin{align}
   \Neq^{++} = \Neq^{--} = \frac{w}{\beta c_s^2}, \qquad
\Neq^{\shat\shat} =  C_p\, , 
\end{align}
which should be compared with 
Eqs.~\eqref{pflucts1}, \eqref{eq:corr_pandsh}, and \eqref{shatflucts} of
the previous section. 
Note that the fluctuations of $\shat$ are uncorrelated
with the pressure fluctuations $\delta x^{(\pm)}$.

%
%
%

%
%

We can now determine the dynamical equation for $N^{\a\b}$ by expressing \Eq{hydro-kinetic} in the eigen-basis of $L$, after defining the matrix
elements
\begin{align}
{\cal D}^{\alpha}_{\,\,\,\, \beta} \equiv e^{(\alpha)}_{\bar a} {\cal D}^{\bar a}_{\,\,\,\,\bar b}\, e_{(\beta)}^{\bar b}\, ,
\end{align}
In the eigen-basis of $L$ the diagonal components $\(L\cdot N-N\cdot L\)^{\a\a}$ vanish, and $N^{\a\a}$ will dissipate but will not oscillate as a function of time. 
By contrast, the off-diagonal components of $\(L\cdot N-N\cdot L\)^{\a\b}$ are
found to be proportional to $c_{s}$, and rotate rapidly. 
This observation allows us to neglect off-diagonal components of $\Nn^{\alpha\beta}$ and to focus on the evolution of $\Nn^{\a\a}$. This 
kinetic (or WKB) approximation to the linearized hydrodynamic wave equations is
described in greater detail in Refs.~\cite{Akamatsu:2016llw,RYZHIK1996327}.
Taking the diagonal components of \Eq{hydro-kinetic}, 
we find: 
\begin{align}
\label{eq:hydrokinetic_equil0}
\partial_t \Nn^{\alpha\alpha}(t,{\bm k})
&= - 2 D_{\a}\, k^{2}
\left(\Nn^{\alpha\alpha}(t,k) - \Neq^{\alpha\alpha}\right)\, , 
\end{align}
where we have used the fact that $N^{\a\b}_{0}$ is a diagonal matrix.
The diffusion coefficients $D_{\a}\equiv {\cal D}^{\a}_{\,\,\,\,\a}$ can be found by explicit calculation:
\begin{eqnarray}
D_{\pm} &=& \frac{1}{2} \left[
\frac{{\lambda_{B}}}{w c_s^2} \left(\frac{\partial p}{\partial n}\right)^2_{e}
+\frac{1}{w}\left(\zeta + \frac{4}{3}\eta\right)
\right] \, , 
\nonumber \\
D_{\shat} &=& \frac{  T\lambda_{B} }{(n T/w)^2 C_p} \, .
\end{eqnarray}
It is useful to define the thermal conductivity $\lambda_T$
with a Franz-Wiedemann type relation   
\st
\l_T \equiv \frac{T\l_B}{(nT/w)^2}  \, , 
\stp
so that
\st
D_{\shat} =  \frac{\lambda_T }{C_p} \label{D-mode} \, .
\stp

\Eq{eq:hydrokinetic_equil0} extends the
hydro-kinetic equations  of a charge-neutral fluid~\cite{Akamatsu:2016llw} to finite baryon density (see also Refs.~\cite{KadanoffMartin,Stephanov:2017ghc}).
The equilibration rate of $N^{\shat\shat}$ is controlled by diffusion coefficient $D_{\shat}$ in \Eq{D-mode}.  Using the thermodynamic relation,
\Eq{CpandCv}, and the definition of the baryon number diffusion
coefficient, $D_B = \lambda_{B}/(\partial n/\partial\mu)_{T}$,
we can relate $D_\shat$ to $D_B$
\st
\label{eq:dshatdb}
D_{\shat} = D_B  \left( \frac{w}{T C_V c_s^2 } \right)\, . 
\stp
The coefficient in parenthesis approaches unity as $n \rightarrow 0$, 
and is never far from unity for the  baryon densities  explored at RHIC.
Thus, $D_\shat$ can be estimated from the baryon diffusion coefficient, $D_B$.

%


So far,
we have derived a kinetic equation \eqref{eq:hydrokinetic_equil0} which describes the evolution of fluctuations around a uniform static background. 
We now sketch the steps needed to extend our analysis to
an expanding hydrodynamic background, referring to the literature for a more complete treatment~\cite{Akamatsu:2016llw}. 
First, we need to take into account that $ \Neq^{\alpha\alpha}(t)$ as well as $D_{\a}(t) $ will in general depend on $t$.
Second, 
we have to introduce gradient terms 
which account for the expansion of the system.
The explicit expression of such  gradient terms is not important for the subsequent discussion.
What is important, though, is that these terms are proportional to $1/\tau_{Q}$, where $\tau_{Q}$ is the expansion rate  we introduced earlier. 
Therefore in an expanding fluid background, 
the hydro-kinetic equation takes the form (schematically)
\begin{eqnarray}
\label{eq:hydrokinetic_equil}
\partial_t \Nn^{\alpha\alpha}(t,{\bm k})
&= - 2 D_{\a}(t)\, k^{2}
\[ \Nn^{\alpha\alpha}(t,k) - \Neq^{\alpha\alpha}(t)\]
+\left[\textrm{terms $\propto 1/\tau_{Q}$}\right]
\, .  
\end{eqnarray}

The dynamics of $N^{\a\a}$ as described by \Eq{eq:hydrokinetic_equil} is driven by the competition between the expansion of the system and the equilibration of thermal fluctuations.
Since the equilibration of $N^{\a\a}$ is achieved by diffusion with rate $\propto Dk^2$,  
$N^{\a\a}$ will depend on non-trivially on wavelength,  although the equilibrium 
expectation $N^{\a\a}_{0}$ is $k$-independent. 
Away from the critical point, 
we can estimate a non-equilibrium length scale, $\ell_{\rm neq} \sim \ell_{\max}$,
which divides the non-equilibrium and equilibrium fluctuations of the system,
characterizing the transition between the two regimes.
Wavelengths longer than $\ell_{\max}$ are too long to equilibrate 
by diffusion over a time $\tau_Q$.
Recalling the introduction, we parametrize the diffusion constant away from the critical point as 
\begin{eqnarray}
\label{D-para}
D_{0}\sim \frac{\ell^{2}_{0}}{\tau_{0}}\, ,
\end{eqnarray}
where  $\tau_{0}$ is the microscopic relaxation time, and $\ell_{0}$ is a microscopic length. 
Equating the diffusion rate of a mode of wavenumber $k \sim 1/\ell_{\rm max}$
with the expansion rate $\sim 1/\tau_Q$ 
\st
D_0 k^2 \sim 1/\tau_{Q}, 
\stp
we obtain \Eq{lmax} as advertised in the introduction. 

As we discuss below, when the system approaches the
critical point, the length scale $\ell_{\rm neq}$ (which
separates the non-equilibrium and equilibrium fluctuations of the system)
will decrease, and a shorter length $\ell_{\kz}$
will replace $\ell_{\rm max}$.

\subsubsection{Evolution of fluctuations in the hydrodynamic regime near a critical point}

Let us now apply the general kinetic equation 
 obtained in the previous section, 
\Eq{eq:hydrokinetic_equil}, to a system passing close to the QCD critical point. 
Because of  criticality two new features emerge which simplify \Eq{eq:hydrokinetic_equil}.
First, since $N^{\a\a}_{0}$ will become singular near a critical point, 
the percent change per time of $N^{\a\a}_{0}$ will become much larger than $1/\tau_{Q}$ (see  below), 
and  the gradient terms proportional to $1/\tau_{Q}$ in \Eq{eq:hydrokinetic_equil} can be safely neglected.  
Second, 
a hierarchy of relaxation rates emerges near a critical point with $D_{\shat}\ll D_{\pm}$~\cite{Stephanov:2017ghc}. 
This is because $D_{\shat}$ is inversely proportional to $C_{p}$,
which is the most divergent 
susceptibility near the critical point.
Thus,  the $\shat$ mode
will be the first to fall out of equilibrium during a transit of the critical point.
For these reasons, 
we will concentrate on the evolution of the $N^{\shat\shat}$ from now on, 
and write the equation for $N^{\shat\shat}$ from \Eq{eq:hydrokinetic_equil} as
\begin{eqnarray}
\label{N-master}
\partial_t \Nn^{\shat\shat}(t,{\bm k})
= -2 D_{\shat}(t) k^2 \, \[\Nn^{\shat\shat}(t,{\bm k})- C_{p}(t)\]\, .
\end{eqnarray}
\Eq{N-master} is valid in the hydrodynamic region $k\ll 1/\xi$. 
We will extend \Eq{N-master} to the scaling regime in the next section. 
%

\subsection{Evolution of fluctuations in the scaling regime near a critical point }
\label{sec:kxi1}


Before continuing, let us review the equilibrium result
for $\Nn^{\shat\shat}$ which is notated with $\Neq^{\shat\shat}$. 
As derived in \Sect{sec:statics},
the equilibrium correlator takes the form
\begin{align}
\label{Css-eq}
\Nss_{0}(t,\k)
   =& \frac{ \Cp(t) } {1 + (k\xi(t))^{2-\eta}}  \, ,
\end{align}
where $C_p(t) = \chi_\is(t)/A_s^2$ and $\xi(t)$ 
are the time dependent susceptibility and correlation length respectively.
The interpolating form for the $k$-dependence 
captures  two limits:
the low-$k$ hydrodynamic limit $k\xi \ll 1$,  and the high-$k$ scaling 
limit $k\xi \gg 1$. 
In the high-$k$ scaling limit, the equilibrium correlation functions 
are power laws $\Neq^{\shat\shat} \propto k^{-(2-\eta)}$ and are independent of $\xi(t)$.

We will introduce a dynamical model to describe the non-equilibrium evolution of $\Nn^{\shat\shat}(t,\k)$  for the full range of momenta, including $k\xi \sim 1$.
Using fluctuating hydrodynamics 
 we derived a hydro-kinetic  equation for  $\Nn^{\shat\shat}$
which applies in the hydrodynamic regime where $k \ll 1/\xi$.
To generalize this relaxation equation to modes in the scaling regime $\xi^{-1} \ll  k \ll
\ell_{0}^{-1}$,  let us first write the small $k$ hydrodynamic
equation~\eqref{N-master} more explicitly 
\begin{align}
\label{Gamma-hydro}
\partial_t \Nn^{\shat\shat}(t,{\bm k})
   =& -2  \left(\frac{\lambda_T }{\Cp}\right) k^2\, \left[ \Nn^{\shat\shat}(t,{\bm k})-  \Cp(t) \right]\,,    & (k \xi \ll& 1) \, .
\end{align}
Here $\lambda_T$ is the thermal conductivity described in \Sect{sec:hydro-kinetic}.
Observe that \Eq{Gamma-hydro} follows the general pattern that the relaxation rate is proportional to the transport coefficient (i.e. $\l_T$) divided by the corresponding susceptibility (i.e. $C_{p}$). 
We expect this pattern will still hold for finite $k$. 
Thus, as a model 
for $k\xi \sim 1$
we will replace  the specific heat in \eqref{Gamma-hydro} with its $k$-dependent
form
 \st
 \Cp  \rightarrow  \frac{\Cp }{1 + (\kb)^{2-\eta}  } \, ,
 \stp
and treat the conductivity $\lambda_T$ as a constant.  
The model takes the
form of a $k$-dependent relaxation time equation
\st
\label{master}
\partial_t \Nn^{\shat\shat}(t,{\bm k})
= -2 \G_{\shat}(t,k) \, \[\Nn^{\shat\shat}(t,{\bm k})- \Neq^{\shat\shat}(t, k)\]\, ,
\stp
where
\st
\label{eq:gammashat}
  \Gamma_{\shat}(t,k)  \equiv \, \left(\frac{\lambda_T}{\Cp \xi^2} \right) (k\xi)^2 (1 + (k\xi)^{2-\eta}) \, .
\stp
The model reduces to the hydrodynamic limit in \eqref{Gamma-hydro}  for $k\xi \ll 1$,  and will approach the universal scaling form for $k\xi \gg  1$.

In the next paragraph we will discuss the limitations of the model,  after describing 
the behavior of relaxation rate $\Gamma_{\shat}$ during a transit of the critical point.
We have already mentioned that for $k \ll \xi^{-1}$, the relaxation rate is small and approaches zero as $k^2$ as is typical of conserved quantities.
We now turn to the relaxation rate at $k\sim \xi^{-1}$ and $k\gg \xi^{-1}$. 
Consider for simplicity the behavior of $\Gamma_\shat$  when the system passes 
directly through the critical point, $t_{\rm cr} \rightarrow 0$ with $t < 0$. 
In this case (from eqs.~\eqref{chitdependence} and \eqref{xitdependence}) the specific heat 
follows the power law
\st
\label{Cp-xi}
   C_{p} =   \frac{\chi_0}{A_s^2} \left(\frac{\xi}{\ell_0}\right)^{2- \eta}  \, ,
\stp
and the relaxation rate for $k\xi=1$ depends on the correlation length $\xi$ as
\st
\label{Gammashatxi}
  \Gamma_{\shat}(t,\xi^{-1})  = \frac{2}{\tau_0} \left(\frac{\xi}{\ell_0}\right)^{-4 + \eta} \, ,
\stp
where we have defined a typical microscopic timescale $\tau_0$ using the
previously defined constants 
\st
\label{tau0-def}
  \frac{1}{\tau_0} \equiv 
 A_s^2 \left(\frac{\lambda_T}{\chi_0 \ell_0^2} \right) \, .
\stp
$\tau_0$ and $\ell_0$ set the diffusion coefficient 
away from the critical point, $D_{0}\equiv\lambda_{T}/C_{p,0}=\ell^{2}_{0}/\tau_{0}$. 
For $k \gg \xi^{-1}$,  the relaxation rate is large, scales with 
 a power of $k$, and is independent of the correlation length
\st
  \label{Gammahigh}
  \Gamma_{\shat}(t, k)  \Big|_{k\xi \gg 1} = \frac{1}{\tau_0} (\ell_0 k)^{4-\eta}   \, .   
\stp

Now let us discuss the limitations of the model. 
In general the conductivity $\l_T$ will scale with the correlation length $\xi$ as
\begin{eqnarray}
\label{lambda-xi}
\l_T &=&\, \l_{T0}\, \(\frac{\xi}{\ell_{0}}\)^{x_{\l}}\, , 
\end{eqnarray}
where $\l_{T 0}$ is the typical thermal conductivity away from the critical point, and the exponent $x_\l$ results from the renormalization of
the conductivity  by the thermal fluctuations of the system.
Such a renormalization (which ultimately is a resummation of the non-linear interactions of the
stochastic system) is neglected in the current model based on linearized
hydrodynamics, which evolves the two point functions and neglects the evolution
of higher point functions.   
Thus the model corresponds to ``model B'' according to the classification of Hohenberg and Halperin~\cite{RevModPhys.49.435}, 
while the dynamical universality class of the QCD critical point is ``model
H''~\cite{Son:2004iv}  where the conductivity is renormalized with  critical exponent
 $x_{\l}=0.946$.
%
In addition, the renormalized conductivity will in general
depend on $k$  as $\l_T K_{\l}(\kb)$, where $K_{\l}(\kb)$ is another dynamical scaling
function with fixed normalization, $K_{\l}(0)=1$. The scaling function
$K_\lambda(k\xi)$ has been studied extensively~\cite{onuki2002phase,Natsuume:2010bs,KAWASAKI19701}, and
its asymptotic behavior is also related to critical exponent $x_{\l}$ 
\begin{eqnarray}
   K_{\l}(\kb) \sim  (\kb)^{-x_{\l}}\, , 
\qquad
\kb \gg 1 \, . 
\end{eqnarray}
Thus, the relaxation rate at large $k$
is generally expected to scale with
the dynamical critical exponent $z\equiv4-\eta-x_\l$ 
\st
\label{Gammahighgood}
    \G_{\shat}(t,k)  \Big|_{\kb \gg 1}  \sim  \frac{1}{\tau_0} (\ell_0 k)^z\, , \qquad z\equiv4-\eta-x_\l \, .
\stp
In comparison with \Eq{Gammahighgood}, the current model \eqref{Gammahigh} has the  dynamical 
critical exponent 
\st
            z = 4- \eta \, ,
\stp
which we will use in the numerical work below.
While it is straightforward to refine the model and  to input $x_{\l}$ and $K_{\l}$ from ``model H'', 
we will continue to use the ``model B'' results, which  are sufficient for the our illustrative purpose. 
It would be interesting to 
simulate a stochastic non-linear Landau-Ginzburg functional  which would  naturally
reproduce the correct dynamical critical exponents of model $H$, and  correctly
describe the  non-linear and non-equilibrium evolution of the system during
the expansion.

\subsection{Kibble-Zurek scaling and missing the critical point} 
\label{sec:KibbleZurek}
Before solving \Eq{eq:gammashat} numerically, let us analyze the timescales 
associated with this evolution.
As noted in the previous subsection, 
low momentum modes with $k \ll \xi^{-1}$ have a small relaxation rate and are out-of-equilibrium even away from the critical point. 
On the other hand, high momentum modes with $k\gg \xi^{-1}$ have a large relaxation rate and  are always in equilibrium.
We will focus on modes with $k \sim \xi^{-1}$ where the relaxation rate as a function of time
follows the pattern  described in \Sect{sec:timescale} for $\chi(t)$ and $\xi(t)$ (see \Eq{eq:gammashat}). 
Specifically,
from eqs.~\eqref{chitdependence}, \eqref{xitdependence}, and \eqref{eq:gammashat},
$\Gamma_\shat$ takes the form 
\begin{eqnarray}
\label{Gamma-rel}
   \G_\shat(t,\xi^{-1})
&=&
   \frac{1}{\tau_{0}} \left(\frac{|t|}{\bar\tau_Q} \right)^{a \nu z} f_\Gamma\left(\frac{t}{|t_\crs|}\right) \, ,
\end{eqnarray}
where  
\begin{eqnarray}
f_\Gamma \equiv  \frac{1}{f_{\chi}\, f^{2}_{\xi}}\, . 
\end{eqnarray}
is a universal function.
Following the pattern described in \Sect{sec:timescale}, $f_{\Gamma}$  has the following limits 
\st
f_\Gamma(u) =   \begin{cases}  1    & u < -1 \,, \\
   f_{\Gamma}^+ \equiv  0.3427 &   u \rightarrow + \infty \, ,
\end{cases}
\stp
and $|u|^{a \nu z} f_{\Gamma}(u)$  is regular for $u>-1$.

Examining the relaxation time equation \eqref{master}, the dynamical evolution of $\Nss(t,k)$ is controlled by a competition between  the relaxation rate $\G_{\shat}(t,k)$ and the rate of change of the equilibrium expectation $\Nss_{0}(t,k)$. 
First we analyze
the limit $t_{\crs} \rightarrow 0$ and $t<0$, where the relaxation rate
takes the scaling form
\begin{align}
   \label{eq:gammapower}
   \Gamma_\shat(t,\xi^{-1}) =&  \frac{1}{\tau_{0}} \left(\frac{|t|}{\bar\tau_Q} \right)^{a \nu z} \, , 
\end{align}
reflecting the equilibrium scaling of $\chi$ and $\xi$ in this limit
\begin{align}
   \label{eq:cppower}
   \Cp=& \frac{\chi_0}{A_s^2} \left(\frac{|t|}{\bar\tau_Q} \right)^{-a \gamma}\, ,  \\
   \xi =& \ell_0 \left(\frac{|t|}{\bar\tau_Q} \right)^{-a \nu}  \,.
\end{align}
For $t>0$ these forms are multiplied by the order one factors, $f_\Gamma^+, f_\chi^+$ and $f_\xi^+$, respectively.
When $t \rightarrow 0$, the system approaches the critical point,  
and the relaxation rate decreases exhibiting critical slowing down.
By contrast, the percent change per time of the equilibrium susceptibility $C_p$  is of order  
\st
\label{Cp-change}
  \left| \frac{\partial_t C_p}{C_p}\right| \sim  \left|\frac{1}{t} \right| \, ,
   \stp
which diverges near the critical point.
Consequently, the system will inescapably fall off equilibrium  at some time $t_{\KZ}$ (the Kibble-Zurek time),
which can be determined by comparing these competing  rates 
\begin{eqnarray}
\label{tKZ-1}
\frac{1}{t_{\KZ}}&=&
\frac{1}{\tau_{0}}\,
   \left(\frac{t_{\KZ}}{\bar\tau_Q}\right)^{a\nu z}\, . 
\end{eqnarray}
Solving for $t_{\KZ}$  we find
\st
t_{\KZ} = \tau_0 \left(\frac{\tau_0}{\bar \tau_Q} \right)^{- a\nu z/(a \nu z + 1) } \, ,
\stp
which is an intermediate scale $\tau_0 \ll t_{\KZ} \ll \bar{\tau}_Q$. 
Indeed, since $\bar \lambda \equiv \tau_0/\bar \tau_Q \ll 1$, the timescales $\tau_0$, $t_{\KZ}$, and $\bar \tau_Q$ are widely separated:
\st
\bar{\lambda } \ll \bar\lambda^{1/(a\nu z+1)} \ll 1 \, .
\stp

The Kibble-Zurek time $t_{\KZ}$ 
characterizes the temporal evolution of $\Nn^{\shat\shat}$ during a transit of the critical point.
Let us introduce an associated length scale $\ell_{\KZ}$ (the Kibble-Zurek
length), which is defined as the value of correlation length $\xi$ at
$t=-t_{\KZ}$
\begin{equation}
\label{lKZs}
   \ell_{\KZ} \equiv \ell_{0} \left( \frac{\tau_0}{\bar{\tau}_Q} \right)^{-a\nu /(a \nu z + 1)}
   =\ell_{0}\, \bar{\l}^{-a\nu /(a \nu z + 1)}\, . 
\end{equation}
Modes with $k \lesssim \ell^{-1}_{\KZ}$ will fall out equilibrium for $|t| \sim  t_{\KZ}$, while modes with $k\gg \ell^{-1}_{\KZ}$ will remain equilibrated. 
We therefore expect that $\ell_{\KZ}$ will characterize the momentum dependence
of $\Nn^{\shat\shat}(t,k)$. $\ell_{kz}$ is also an intermediate scale, $\ell_0
\ll \ell_{\KZ} \ll \ell_{\rm max}\sim \ell_{0}\, \l^{-1/2}$, where $\ell_{\rm
max}$  is the maximum wavelength
that can be equilibrated  away from the critical point.  

 Finally, since  the evolution is ``frozen'' for $t \gtrsim  {-}t_{\KZ}$, 
 the magnitude of $\Nn^{\shat\shat}(t,k)$ 
can be estimated by the value of $C_{p}$ at $t=-t_{\KZ}$
\begin{align}
 \label{chi-KZ}
 \Nn^{\shat\shat} \sim 
 \frac{\chi_{\KZ}}{A_s^{2}}\equiv C_{p,\KZ}\, ,  \qquad   \chi_{\KZ} \equiv \chi_0 \left(\frac{\ell_{\KZ}}{\ell_0}\right)^{2-\eta}\, .
 \end{align}
Thus, $\ell_{\KZ}$ also determines the magnitude of  fluctuations during a
transit of the critical point through the definition of $\chi_\KZ \propto \ell_{\KZ}^{2-\eta}$.

The qualitative discussion in the preceding paragraphs motivates us to introduce a rescaled two point function
\begin{eqnarray}
\label{N-scale1}
\Nn^{\shat\shat} 
\equiv
   \frac{\chi_{\kz}}{A_s^{2}}
\, \, \Nssbar(\tb, k \lKZ ;t/|t_{\cross}|)\, ,
\end{eqnarray}
where we anticipate $\Nssbar$ will be of order unity, and will depend on the rescaled time
\st
\label{scaling-variable}
\tb \equiv \frac{t}{t_{\kz}}\, . 
\stp
Substituting \eqref{N-scale1} into \eqref{master}, 
we obtain an equation for $\Nssbar$:

\begin{subequations}
\label{master0}
   \begin{align}
\pd_{\tb}\, \Nssbar=&
- 2\, 
      |\tb|^{a\nu z} \, \frac{(k\xi)^2}{K_{\chi}(\kb)}\,
      \left[ \Nssbar - |\tb|^{-a\g}\, K_{\chi}(\kb)\right]
      &
      \tb\leq& \blue{-}t_\cross/t_\KZ \, ,  \\
\pd_{\tb}\, \Nssbar=&
- 2\, 
      |\tb|^{a\nu z} \, f_\Gamma \, \frac{(k\xi)^2}{K_{\chi}(\kb)}\,
      \left[ \Nssbar - |\tb|^{-a\g}\, f_\chi \, K_{\chi}(\kb)\right]
      &
      \tb\geq& \blue{-}t_\cross/t_\KZ \, ,
   \end{align}
where
\st
  k \xi = \begin{cases} k\ell_{\KZ} \;  |\tb|^{-a\nu } 
       &       \bar{t}  \leq  \blue{-}t_\cross/t_\KZ \\
     k\ell_{\KZ} \;  |\tb|^{-a\nu }    f_{\xi} &   \bar{t} \geq  
     \blue{-}t_\cross/t_\KZ  
  \end{cases} \, ,
  \stp
  and  $K_{\chi}$ is given \Eq{eq:OZform}. 
The three scaling functions $f_\Gamma$, $f_\chi$ and $f_\xi$ take the form
  \st
    f_{\Gamma}\left(\frac{t_\KZ}{|t_\cross|}\, \tb\right),
    \qquad    f_{\chi}\left(\frac{t_\KZ}{|t_\cross|}\, \tb\right),
\qquad    f_{\xi}\left(\frac{t_\KZ}{|t_\cross|}\, \tb\right)\, . 
  \stp
\end{subequations}
 Thus, $\bar{N}^{\shat\shat}$ only depends on scaling variables $\bar{t}=t/t_{\KZ}, k \lKZ$ and $t_{\KZ}/\tcross$. 
 When the system passes directly through the critical point $t_{\cross} \rightarrow 0$, the quantities $f_\Gamma$, $f_\chi$ and $f_\xi$ approach universal constants ($f_\Gamma^+$, $f_\chi^+$, and $f_\xi^+$), and the correlation function $\Nn^{\shat\shat}$ is only
a function of $t/t_{\KZ}$ and $k\, \ell_{\KZ}$.
When the system misses the 
critical point by an amount $\Delta_s$  there is an additional time scale $t_\cross \propto \tau_Q \Delta_s^b$, 
%
and the correlation function for $t > {-}t_\cross$ additionally depends on the ratio  $t_{\cross}/t_\KZ$. 
We will present numerical results for  $\Nssbar$ in the next section by solving \Eq{master0}.

\subsection{Transits of a critical point: numerical evaluation }
\label{sec:evaluation}

\begingroup
\centering
\begin{figure}
\centering
\subfigure[\;Before the critical point]{\includegraphics[width=0.45\textwidth]{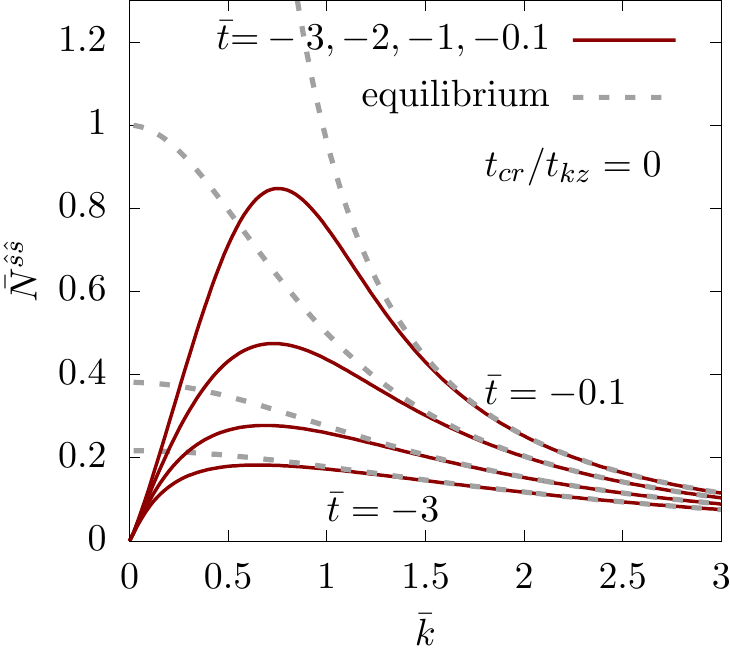}}\label{fig:11a}
\subfigure[\;After the critical point]{\includegraphics[width=0.45\textwidth]{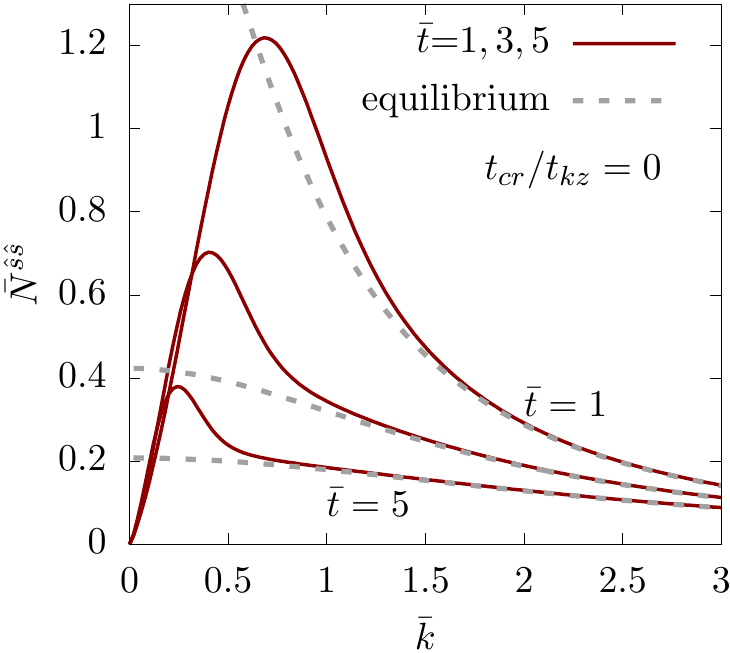}}\label{fig:11b}
   \caption{
   \label{fig:tcross0} 
      The time evolution of the correlation function of entropy per baryon fluctuations
   $\delta \hat s\equiv \delta s - (s/n) \delta n$, when the system passes
   directly through the critical point, $t_{\cross}/t_{\KZ}=0$. 
   The wavenumber is measured
   in units of $\ell_{\KZ}^{-1}$ ($\bar{k}\equiv k \ell_{\KZ}$),  and $N^{\shat\shat}$ has been rescaled by the specific heat $C_p$ at the Kibble-Zurek time $C_{p, \KZ}$ ($\bar{N}^{\shat\shat}\equiv\Nss/C_{p,\KZ}$) 
   (a) The time evolution in the coexistence region $t<0$ (before the critical point).
   (b) The time evolution after the system has left the coexistence region $t>0$ (after the critical point). 
}
\end{figure}
\endgroup
%
%

Now we will determine $\Nss$ by solving \Eq{master0} numerically\footnote{
We need to specify the initial conditions of $\Nss$ at an initial time $t_{I}/t_{\KZ}$, 
where $t_{I}<0$ is the time when system enters the critical region. 
However, we are working in the parametric regime where  $\tau_{Q}/t_\KZ \to \infty$, 
and the time $t_{I}$ is  of order $\tau_{Q}$.  Therefore,  $t_{I}/t_{\KZ}$ should be taken 
to negative infinity;   we take  $t_{I}/t_{\KZ}\sim -40$ in practice.
Non-equilibrium effects will not be important at this early time, 
and consequently we initialize $\Nss$ in equilibrium.}. %
First, we evaluate $\Nss$    when the system passes directly through the critical
point by setting $t_{\rm cross}$ to zero.
In \Fig{fig:tcross0}(a) and (b) we plot $\bar\Nn^{\shat\shat}$ as function of $\bar k$ for representative times before and after the critical point respectively.
For comparison, we plot the corresponding equilibrium expectation with dashed curves. 

As seen in the figure,  the fluctuations  recorded by $\Nss$ are maximal at a given wavenumber
$k_{\neql}$
corresponding to a definite wavelength,
   $\ell_{\neql}\equiv k_{\neql}^{-1} \sim \ell_{\KZ}$. 
This is in contrast with the behavior of the equilibrium fluctuations $\Nss_{0}$ (the dashed curves) 
which increase monotonically as  $k \rightarrow  0$. 
The maximum is the result of  a competition between the   hydrodynamic behavior at small $k$, and the critical scaling behavior at large $k$. 
Modes with $k \ll \ell_{\neql}^{-1}$  equilibrate slowly (diffusively), reflecting the fact that the total charge  is conserved and does not fluctuate. 
Consequently, 
the system does not respond to the increasing critical susceptibility  at small $k$, and the magnitude of $\Nss$ in the hydrodynamic region remains small compared to the equilibrium specific heat $C_{p}$. 
At large $k$, the relaxation rate grows as $k^{z}$ 
and becomes very large.
Thus, the large $k$ tail of $\Nss$ is always close to the equilibrium expectation, which vanishes as $1/k^{2-\eta}$. 
To summarize, $\Nss$ will become small at both small $k$ and large $k$,
naturally exhibiting maximum at some intermediate wavenumber $\ell_{\neql}^{-1}$.
This scale characterizes $\Nss$ in the sense that wavenumbers significantly larger than $\ell_{\neql}^{-1}$ are in equilibrium, while those smaller than $\ell_{\neql}^{-1}$ are out of equilibrium. 

From Fig.~\ref{fig:tcross0}, the fluctuations grow with time for $t<0$, and then return to their typical size after passing the critical point, $t>0$.
However, as we approach the critical point, the growth in $\Nss$ for $t >
-\tKZ$ is  modest when compared to the rapid growth of $C_{p}$ (the dashed
curves at $k=0$). The system is exhibiting critical slowing down, and
lags behind its equilibrium expectation.

The slow evolution of $\Nss$ implies that the system can remember the magnitude of the critical fluctuations even after passing through the critical point.
Indeed  for $\tb>0$, $\Nss_{\max}$ is even larger than its equilibrium expectation. 
Similar observations about the ``memory effect'' of critical fluctuations have been made in previous studies~\cite{Mukherjee:2015swa,Berdnikov:1999ph}.
The distinctive feature of $\Nss$, namely the maximum at a specific wavenumber $\ell_{\neql}^{-1}$, 
is remembered for $\tb>0$.
It remains to be seen which experimental observables provide access to this interesting structure
-- see Sec.~\ref{sec:exp} for a preliminary 
proposal.


%
%
\begingroup
\centering
\begin{figure}
%
%
\subfigure[\;Inside the coexistence region]{\includegraphics[width=0.45\textwidth]{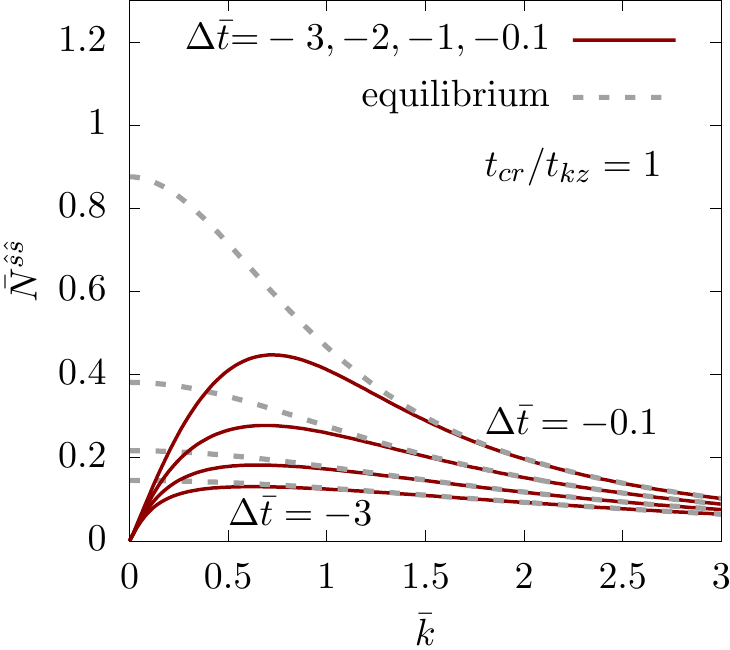}}
\subfigure[\;Outside the coexistence region]{\includegraphics[width=0.45\textwidth]{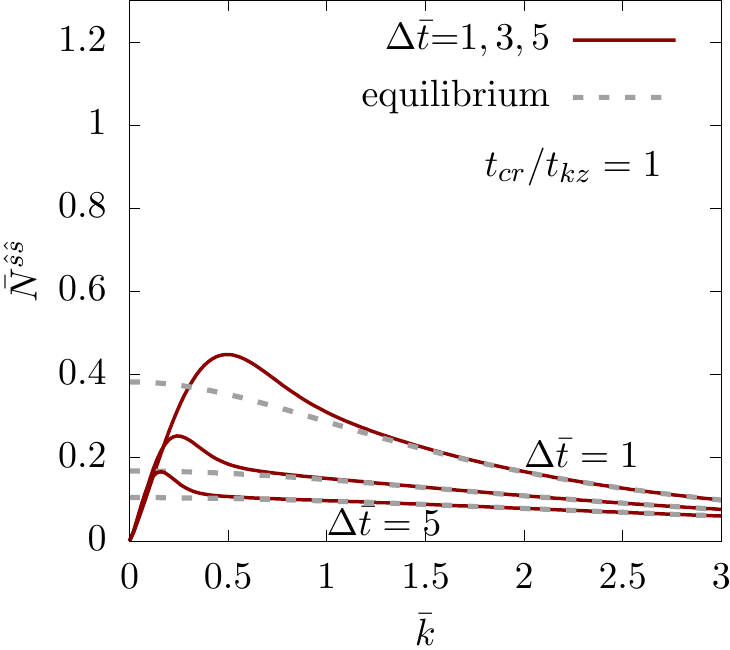}}
\subfigure[\;Inside the coexistence region]{\includegraphics[width=0.45\textwidth]{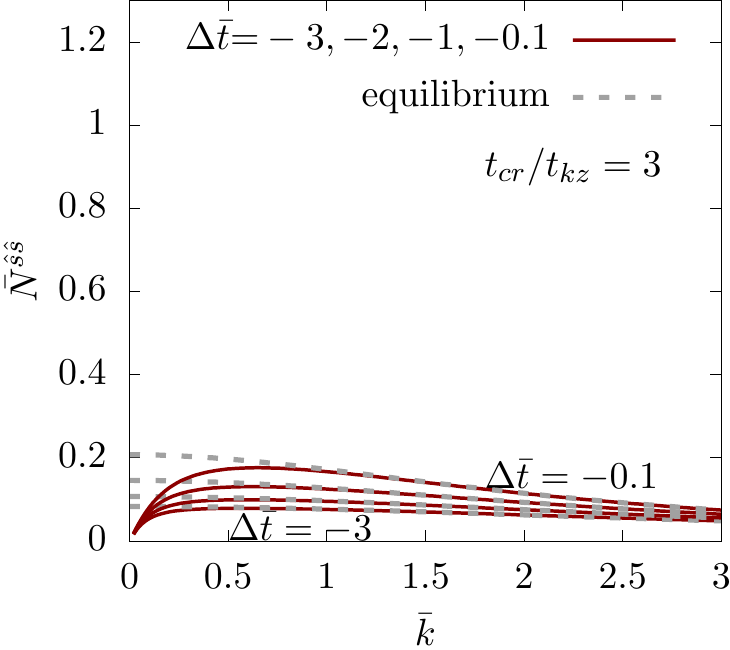}}
\subfigure[\;Outside the coexistence region]{\includegraphics[width=0.45\textwidth]{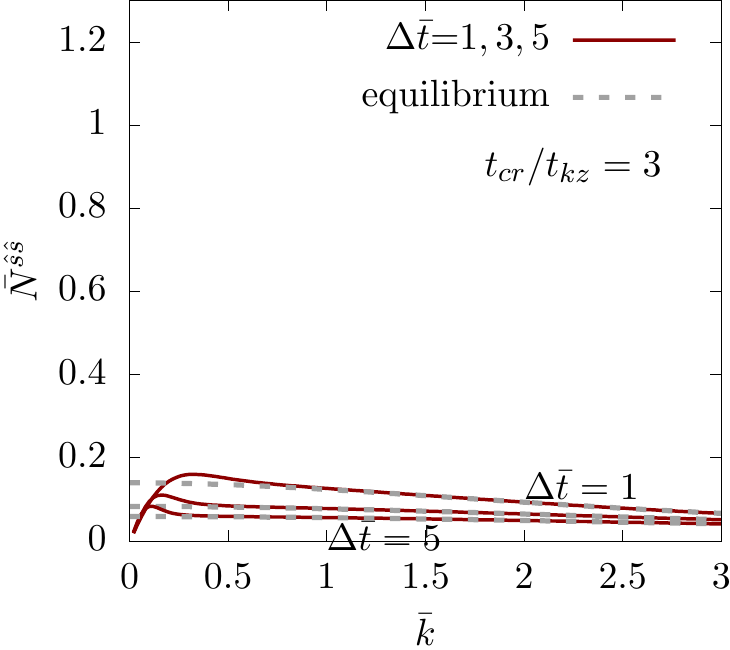}}
   \caption{
   \label{fig:tcross1} 
(The upper row):
The same as \Fig{fig:tcross0},  but the system misses the critical point with $t_\cross/{t_{\KZ}}=1$. 
  $t_\cross$ is the time when the  system leaves the coexistence region, and  $\Delta t\equiv t - t_\cross$.
  (a) The time evolution of $N^{\shat\shat}$  when the system is in the coexistence region $t < t_\cross$, and (b)
   when the system leaves the coexistence region $t > t_{\cross}$.  
   (The lower row):
  The same as the upper row,  but with $t_\cross/{t_{\KZ}}=3$.
}
\end{figure}
\endgroup

We now turn to finite detuning case shown in Fig.~\ref{fig:tcross1}.
In Fig.~\ref{fig:tcross1} (a,b), 
we show our results for $N^{\shat\shat}$ at $t_{{\rm cr}}/t_{\KZ}{=}1$.  
The qualitative features are similar to the $t_{{\rm cr}}/t_{\KZ}{=}0$ case,
but the magnitude of the fluctuations is reduced.
For still larger detuning $t_\cross/t_\KZ{=}3$ shown in (c,d), the fluctuations
are reduced even further.  In the  large detuning regime 
the equilibrium scaling of the specific heat at the crossing time $t_{\cross}$
determines the magnitude of the fluctuations rather than the relaxation
dynamics.  Thus, the magnitude of the critical fluctuations are independent of
$\lambda$ in this regime.
Straightforward analysis based the previous sections (see \Sect{sec:timescale}) shows
that at $t_{\cross}$ the fluctuations are of order
\st
N^{\shat\shat}  \sim C_{p,\KZ} \left( \frac{t_{\cross}  }{t_{\KZ}} \right)^{-a\gamma}  \sim \frac{\chi_0}{A_s^2}  \, \bar{\Delta}_s^{-\gamma/\beta}  \, .
\stp
The wavenumber $\ell_{\neql}^{-1} $ where the system transitions from the
non-equilibrium behavior at small $k$ to equilibrium behavior at large $k$ is also reduced 
relative to $\ell_{\KZ}^{-1}$.
Equating the relaxation rate at the crossing time 
to the rate of change in  equilibrium, $\Gamma(t_\cross, k_\neql) \sim t_\cross^{-1}$,  shows that 
\st
\ell_{\neql} \sim  \ell_{\KZ} \left( \frac{t_{\cross}}{t_{\KZ}} \right)^{(a\nu (z -2) +1)/2}   \sim \ell_0\,  \lambda^{-1/2}\, \bar{\Delta}_s^{(a \nu (z - 2) + 1)/2 a\beta}  \, .
\stp
Numerically these exponents evaluate to 
\begin{gather}
N^{\shat\shat}  \sim \frac{\chi_0}{A_s^2}  \, \bar{\Delta}_s^{-3.8}\, ,  \\
\ell_{\neql} \sim  \ell_0 \lambda^{-1/2} \, \bar{\Delta}_s^{3.26} \, .
\end{gather}
When the detuning $\Delta_s$ approaches unity, the non-equilibrium length
$\ell_{\neql}$ approaches $\ell_{\rm max} = \ell_0 \lambda^{-1/2}$. Modes with
wavelength longer than $\ell_{\max}$ remember the initial conditions at
$t{=}-\tau_Q$, and are  unaffected by the transit of the critical point.

Summarizing this subsection, 
we have evaluated the fluctuations in the entropy to
baryon number, $\Nss$, for a system which passes directly through the critical point ($t_{\cross}/t_{\KZ}{=}0$), and which misses the critical point ($t_{\cross}/t_{\KZ}{\notequal} 0$). 
When $t_{\cross}/t_{\KZ}$ is not significantly larger than unity, 
the wavenumber dependence of $\Nss$ is qualitatively different from its equilibrium expectation, and from earlier work.
Previously, the non-equilibrium variance of the order parameter field has been evaluated for ``model A''~\cite{PhysRevB.86.064304}.
In this case the order parameter is not conserved, and  its relaxation rate remains finite at $k=0$. 
By contrast, the order parameter for QCD is conserved, and the relaxation rate vanishes as $k\rightarrow 0$. 
Because of this fundamental difference
$\Nss$ develops a maximum around $k\sim \ell_{\KZ}^{-1}$, 
and the critical fluctuations will be most pronounced at the
corresponding wavelength $\sim \ell_\KZ$.
This feature is absent in the study of Ref.~\cite{PhysRevB.86.064304}. 

\section{Discussion
\label{sec:discussion}
}

\begin{figure}
   \begin{center}
\includegraphics[width=0.47\textwidth]{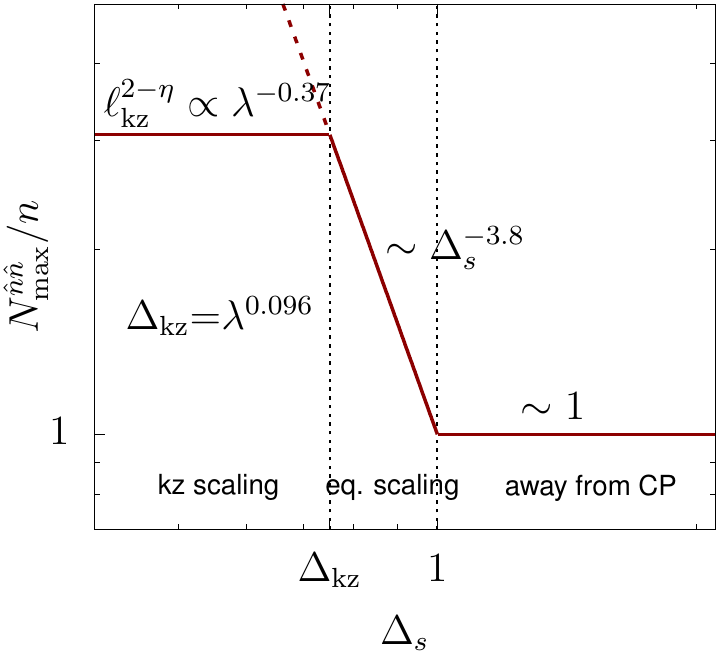}
\hspace{0.1in}
\includegraphics[width=0.47\textwidth]{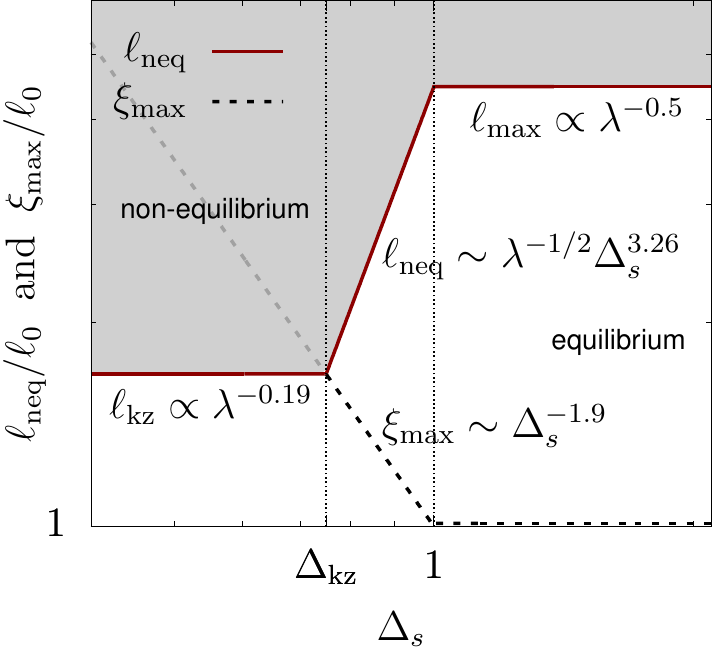}
\end{center}
   \caption{\label{fig:summary}
      A schematic plot showing the  dependence of the maximal 
      fluctuations, $N^{\hat n\hat n}_{\rm max}/n$, and the maximal correlation length,
      $\xi_{\rm max}$, on the parameters $\Delta_{s}$ and $\lambda$ during a transit of the critical point. Also shown is the non-equilibrium length
      $\ell_{\neql}$ --  modes with wavelength longer than $\ell_{\neql}$ fall out
      of equilibrium during the transit.
For $\Delta_{s} \gtrsim 1$ the adiabatic trajectory misses the critical point completely. In this regime
$N^{\hat n \hat n}_{\rm max}/n$ is of order unity, and $\ell_\neql$ is of order $\ell_{\rm max} \sim \ell_0 \l^{-1/2}$. 
For $\Delta_{s} <1$,  but  larger than $\Delta_{\KZ}{=}\lambda^{0.096}$, the trajectory approaches the critical point.
In this (narrow) regime the dependence of $\Nnn_{\rm max}/n$ and $\xi_{\rm max}$ on $\Delta_s$ follows from equilibrium scaling. The non-equilibrium length $\ell_{\rm neq}$ remains longer than 
the correlation length $\xi_{\rm max}$. 
For $\Delta_{s} \lesssim \Delta_{\KZ}$, equilibrium scaling is irrelevant, and the Kibble-Zurek scaling sets in. 
In Kibble-Zurek region $\Nnn_{\rm max}/n$ and $\ell_{\neql}$ scale as $\l^{-0.37} $ and $\ell_0\l^{-0.19}$ respectively. In this region, both quantities are independent $\Delta_s$, i.e. of how close the adiabatic 
trajectory is to the critical point.
}
\end{figure}

In this  paper we have studied how the QCD medium created in a heavy ion collision 
will evolve  during a transit of the conjectured critical point.  We have defined two parameters, which
are repeated here for convenience.
The first is the ``detuning parameter"
\st
    \Delta_s
 \equiv   \frac{n_c}{s_c} 
\left( \frac{s}{n} 
- \frac{s_c}{n_c} \right) \,, 
\stp
and the second is the  ratio of the microscopic time scale $\tau_0$ to
the expansion rate $\tau_Q^{-1}$
\st
     \lambda \equiv \tau_0 \partial_{\mu} u^{\mu} \equiv \frac{\tau_0}{\tau_Q} \, .
\stp
Then we asked how the critical hydrodynamic fluctuations in the system depend on these two parameters during
the transit.   
These two parameters quantify how missing the critical point and finite relaxation rates
will regulate the growth of critical fluctuations. This conclusion is organized around explaining
\Fig{fig:summary} which summarizes our results.



\subsubsection{Object of study and its behavior away from the critical point}
First we explained that observable of primary interest  is the fluctuations
in the entropy per baryon (multiplied by $n$)
\st
  \hat{s} \equiv n \delta \left(\frac{s}{n} \right) = \delta s - \frac{s}{n} \delta n \, .
\stp
From an experimental point of view, it may be easier to work with the fluctuations in the baryon number per entropy (multiplied by $s$) :
\st
\delta \hat{n}  \equiv s \delta \left(\frac{n}{s} \right) = \delta n - \frac{n}{s} \delta s \, .
\stp
which contains the same physical content. 

   There are several reasons (discussed in \Sect{sec:mappingqcdtoising} and \Sect{sec:hydro-kinetic}) why
$\delta \nh$ is the relevant quantity.
First, $\delta \nh$ is an eigenmode of linearized
hydrodynamics,  and its fluctuations are proportional to the specific heat
at constant pressure. 
Specifically,  the two point functions are defined as
\st
\label{Nnhatnhat}
\Nn^{\hat n \hat n}(t,\k) \equiv \int d^3x \, e^{i \k \cdot \vx}\,  \llangle \delta \hat n(t,\vx) \delta \hat n(t,{\bm 0}) \rrangle \, ,
\stp
and in equilibrium  $\Nn^{\hat n \hat n}$  determines
$C_p$ from its  small $k$ limit
\st
\left. \Neq^{\hat n \hat n}(t,{\bm 0}) \right|_{\rm eq} = \left. V \llangle (\delta \hat n)^2 \rrangle  \right|_{\rm eq} = \left(\frac{n}{s} \right)^2 C_p \, .
  \stp
  In the body of the text we have worked with $N^{\shat \shat}(t,\k)$  which 
  is proportional  $N^{\nh\nh}(t,\k)$ 
\st
\Nn^{\nh \nh}(t,\k) = \left(\frac{n}{s}  \right)^2 \Nn^{\shat \shat}(t,\k) \, .
\stp
As the temperature approaches its critical value, $(n/s)^2 C_p$ will always diverge 
with the largest critical exponent of the Ising
susceptibility matrix, $\gamma \simeq 1.23$. By contrast, the squared speed of sound approaches zero with the  critical exponent $\alpha\simeq0.11$, which is too slow to be of practical interest for the heavy ion program.  As discussed in \Sect{sec:specificheats}, these
statements about $C_p$ and $c_s^2$  are independent  of the detailed  mapping matrix  between the QCD and Ising variables. 

Now let us desribe the behavior of $\Nnn$ away from the critical point 
as illustrated in \Fig{fig:summary}.
Away from the critical point, the fluctuations in $\delta\hat{n}$ 
scale as the fluctuations  in $\delta n$, which can be reasonably 
expected  to be roughly Poissonian, $V \llangle (\delta n)^2 \rrangle \sim n$.  
This leads to a Poisson estimate for these
fluctuations\footnote{For example,  we may estimate $N^{\hat n \hat n}$ for a hadron gas. For a hadron gas at a temperature  of
   $T\simeq 155$ and $s/n\simeq 25$ (corresponding 
   to the chemical freezeout conditions at $\sqrt{s_{\scriptscriptstyle NN}}=12.5\,{\rm GeV}$)
we find $n C_p/s^2 = 0.65$. }
\st
\label{Poissonbase}
\frac{ \Nn^{\hat n \hat n}(t,\k)}{n}  \sim 1 \,, \quad \mbox{with} \qquad   \ell_{\rm max}^{-1} \ll k \ll \ell_0^{-1} \, .
\stp
Here $\ell_0$ denotes a typical microscopic length scale, and $\ell_{\rm max}$ is discussed below.
Searches for critical fluctuations will look for enhancements 
at fixed $k$ to this baseline expectation that change non-monotonically with the (mean) $n/s$.

Note that the  Poissonian expectation in \Eq{Poissonbase}
is independent of $k$  for all equilibrated modes with wavenumber smaller than the inverse correlation length $\sim \ell_0^{-1}$.   
As discussed in the introduction,  modes with wavelength longer than a 
``non-equilibrium'' length, $\ell_{\neql} \sim \ell_{\rm max} \sim \ell_0 \lambda^{-1/2}$, are
always out of equilibrium even away from the critical point~\cite{Akamatsu:2016llw}, and
will not show critical behavior.
We will see that when the system approaches the critical point,
  modes with wavelength shorter than $\ell_{\rm max}$  will begin to fall out  of equilibrium, and the non-equilibrium length $\ell_{\neql}$ will
decrease.
This shown by the grey region  of  \Fig{fig:summary}(b).



\subsubsection{How  missing the critical point regulates the critical fluctuations}

In \Sect{sec:timescale} we determined how the equilibrium susceptibilities in QCD
are regulated in time as the medium passes close the critical point during an adiabatic expansion
with a detuning parameter $\Delta_s$.
This
time evolution follows  a specific pattern, which is a reflection 
of the scaling of the equilibrium equation of state.
For  example, the equilibrium specific heat
$(n/s)^2 C_{p}$ (which diverges like the Ising susceptibility  $ \chi_{\rm is} \propto
r^{-\gamma}$) has the following  time dependence for an adiabatic trajectory
near the critical point
\begin{align}
\label{eq:crossingeq}
\left. \Neq^{\hat n \hat n}(t,{\bm 0}) \right|_{\rm eq}  
=& c_0   \, n \left| \frac{t}{\tau_Q}
\right|^{-a \gamma} f_{\chi}\left( \frac{t}{|\tcross|} \right)   \,.
\end{align}
Here $f_{\chi}(t/|\tcross|)$ is a known universal  scaling function of order unity which  can
be determined by the $(R,\theta)$ parameterization of the Ising Model susceptibility.
$c_0$ is a dimensionless and order
one non-universal constant,  and the ``crossing time" is
\st
\label{tcross:conclude}
t_{\cross}  \equiv -c_1  \tau_Q \Delta_{s}^{1/a\beta} <0\, ,
\stp
where  $c_1$ is another (dimensionless and order one) non-universal constant\footnote{Explicit expressions for these constants are given  in the text ($c_0 =0.365\, A_n^{-a\gamma}$ and $c_1 = A_s^b/A_n$)
   in terms of the mapping matrix $M^{A}_{\sp b}$ between the QCD and Ising variables described in \Sect{sec:map}.}. 
$t/|\tcross|$ plays the role of the scaling variable,  
and the scaling function $f_{\chi}(t/|\tcross|)$ approaches 
a (universal) constant for $t/|\tcross|  \rightarrow \pm \infty$.
From \Eq{eq:crossingeq} we see
that the specific heat grows  like a power
until the scaling variable $t/|\tcross|$ approaches $-1$. For $t/|\tcross| \sim -1$
the system becomes aware that adiabatic trajectory will miss the critical point by 
$\Delta_s$, and this stops the growth of the specific heat.
Setting $t$ to $\tcross$, we can estimate the maximum magnitude of  equilibrium
critical fluctuations relative  to  the Poissonian expectation 
\st
\label{eq:equilibrium_estimate}
 \frac{ \Nn^{\hat n\hat n}_0(t,{\bm k}) }{n}  \sim \Delta_s^{-\gamma/\beta}  \, ,
\stp
Here the wavelengths of interest  $k^{-1}$ are of order the correlation length at the crossing time
\st
k^{-1}  \sim \xi(t_\cross) \sim  \ell_0 \, {\Delta}_s^{-\nu/\beta} \,.
\stp
Sufficiently long wavelength modes are always out of equilibrium and will not show 
the enhancement in \Eq{eq:equilibrium_estimate}. \Sect{sec:evaluation} estimates
that for $\Delta_s$ small (but larger than a $\Delta_{\KZ}$ discussed below) the non-equilibrium length  is of order $\ell_{\neql} \sim \ell_0 \lambda^{-1/2} \Delta_s^{3.26}$. 
\Fig{fig:summary} shows how the correlation length $\xi(t_{\cross})$ and the non-equilibrium length $\ell_{\neql}$ come together as we begin to approach the critical point.


The estimate in \Eq{eq:equilibrium_estimate} realizes one of the goals of this paper, i.e. to parametrically estimate how
missing the critical point limits the critical fluctuations.
However, the analysis in the next section
shows (unfortunately) that non-equilibrium physics will set
in well before  the critical fluctuations are regulated by
a finite missing parameter $\Delta_s$.
Thus, the non-equilibrium dynamics will regulate the critical
fluctuations  well below the equilibrium estimate in \Eq{eq:equilibrium_estimate}.  
For this reason we will refrain from substituting numbers into \Eq{eq:equilibrium_estimate}.

\subsubsection{How critical slowing down regulates the critical fluctuations}

In \Sect{sec:dynamics} we estimated how the finite relaxation time limits 
the growth of critical fluctuations. For conserved (or approximately
conserved) quantities such as $n/s$  the  relaxation time 
depends on the wavelength of the mode of interest, with 
longer wavelengths modes taking longer to relax.  
For  $k \sim \xi^{-1} $, the typical relaxation time increases near
the critical point as
\st
\tau_{R}(\xi) \equiv  \tau_0 \left(\frac{\xi}{\ell_0}\right)^{z}   \, ,
\stp
where $z\equiv 4 - \eta\simeq 4$ in our setup\footnote{
We have defined $\tau_R(\xi) \equiv 1/\Gamma_\shat(t,\xi^{-1})$
used in the body of the text, e.g. \Eq{Gammashatxi}.
   The dynamical 
exponent $z=4-\eta$ is modified to $z=3-\eta$ in a more refined
treatment where the conductivity ${\lambda_B}$ is renormalized by critical
fluctuations.}, and $\tau_0$ is the microscopic time.
We then find that modes with $k \sim \xi^{-1}$ fall  out of 
equilibrium  at the Kibble-Zurek time
\st
\label{tkz:conclude}
t_{\rm kz}  \sim  \tau_0 \left( \frac{\tau_0}{\tau_Q} \right)^{-a\nu z/(a\nu z + 1)}\,,  \qquad \frac{a \nu z}{a\nu z + 1} \simeq 0.74\, ,
\stp
where $\nu \simeq 0.63$. 
The correlation length at this time is 
\st
\ell_{\rm kz} \sim \ell_0 \left( \frac{\tau_0}{\tau_Q} \right)^{ -a\nu/(a\nu z+1)}\,,    \qquad \frac{a\nu}{a\nu z+1} \simeq 0.19 \, .
\stp

Let us compare the $\tkz$ and $\tcross$ timescales. The Kibble-Zurek
dynamics will begin to regulate the growth of critical fluctuations before the 
scaling behavior of the equation of state whenever $\tkz \gg \tcross$.  In
this limit  $\Delta_s \rightarrow 0$ and   the scaling structure of the equation of state
is irrelevant, since the system falls out of equilibrium before reaching
the detailed scaling regime.
Comparing  \Eq{tkz:conclude} and \Eq{tcross:conclude} we
see that  $\tkz \gg \tcross$ whenever $\Delta_s$ is less than a certain threshold $\Delta_\KZ$
\st
\label{t0-tQ-Delta}
\Delta_s  <  \Delta_{\KZ} \equiv \lambda^{a\beta/(a \nu z+ 1) } \,.
\stp
As shown in Fig.~\ref{fig:summary}, for $\Delta_s < \Delta_{\KZ}$ the non-equilibrium length  is set by  $\ell_{\kz}$ and the magnitude of the fluctuations is of order the equilibrium susceptibility  at $t_{\KZ}$. 
Substituting numbers, with $a\simeq 1.12$, $z\simeq 3.96$,  and $\beta=0.32$, 
we find 
\st
\Delta_{\KZ} =  0.86 \, \left( \frac{\l}{0.2} \right)^{0.096}  \, .
\stp
Clearly the strikingly  small power, $0.096$, makes the value $\Delta_{\KZ}$ remarkably insensitive to the value of $\l$.
Thus, for  realistic heavy-ion collisions  with a finite $\lambda$, 
the detailed equilibrium scaling of the equation of state has a  limited range of validity, $\Delta_{\KZ} \ll  \Delta_s \ll  1$.
Essentially, if one is close enough to the critical point, then the dynamics 
will always be out of equilibrium.  Thus,
to simulate the evolution of  trajectories with $\Delta_{s} < \Delta_{\KZ}$, inputting an equation of state with the detailed scaling behavior  (see Ref.~\cite{Parotto:2018pwx}) into the hydrodynamic codes
is not really necessary or sufficient.  
It is essential to simulate the non-equilibrium evolution of the system, along the lines of this work and Ref.~\cite{Stephanov:2017ghc}. 

Let us estimate the Kibble-Zurek timescale.
We have defined a small parameter $\lambda$, and the three time scales in our problem,
\st
     \tau_0  \ll \tkz \ll \tau_Q \, ,
\stp
are of relative size
\st
\tau_0  \ll \tau_0 \,\lambda^{-0.74}  \ll \tau_0 \,\lambda^{-1} \, .
\stp
Taking\footnote{We have
   estimated the hadron density 
   below using a thermal model. Then we 
   multiplied the distance by the typical 
   quasi particle velocity
$\sqrt{3 c_s^2}$  to arrive at this estimate.} $\tau_0\simeq 1.8\,{\rm fm}$ and $\lambda = 0.2$, we find
a relatively long time for $\tkz$ :
\st
1.8\,{\rm fm} \ll 5.8\,{\rm fm} \ll 8.9\,{\rm fm} \, .
\stp
Thus, if the system freezes out over a time of $\tkz \sim 5.8\,{\rm fm}$, then the critical enhancement
of fluctuations estimated below may be visible.

Similarly, the system has 
the length scales 
\st
\label{threelengths}
\ell_0 \ll \ell_{kz} \ll  \ellmax \, ,
\stp
which are of relative size 
\st
\ell_0  \ll \ell_0 \, \lambda^{-0.18} \ll  \ell_0 \, \lambda^{-1/2}  \, .
\stp
The microscopic length $\ell_0$ is of order the inter-particle spacing.
For a hadronic gas  with $n/s=25$  and
a chemical freezeout temperature $T\simeq 155\, {\rm MeV}$  
this length is
approximately, $\ell_0 \simeq 1.2\, {\rm fm}$. Taking $\lambda = 0.2$ 
we find that the three length scales are  of order
\st
1.2\,{\rm fm} \ll 1.6 \, {\rm fm} \ll 2.7\,{\rm fm} \, .
\stp
Comparing these numbers, we see that the correlation length at 
freezeout is at most twice the inter-particle spacing at these
low densities.

Let us estimate the magnitude of the critical fluctuations 
when the Kibble-Zurek dynamics regulates the 
growth.  The timescales and length-scales are set by the Kibble-Zurek time and length.
Substituting $\tkz$ from \Eq{tkz:conclude} into  \Eq{eq:crossingeq}  
(with  $c_0 \sim f_\chi \sim 1$), we find that the  magnitude of the fluctuations 
relative to our Poisson expectation are enhanced by
\st
\label{Nmax-in-l}
\left. \frac{\Nn^{\hat n \hat n}({t_{\rm kz}}, {\bm k}) }{n}  \right|_{k\sim \ell_{\KZ}^{-1}}
   \sim    \lambda^{-\gamma a/(a\nu z + 1) } \, .
\stp
Numerically for $\lambda=0.2$ we find a somewhat anemic 80\%
enhancement 
\st
\left. \frac{\Nn^{\hat n \hat n}(\tkz, {\bm k}) }{n}  \right|_{k\sim \ell^{-1}_{\KZ}}
   \sim   1.8 \left( \frac{\lambda}{0.2} \right)^{-0.37} \, .
\stp
This enhancement $\propto \lambda^{-0.37}$ is illustrated in \Fig{fig:summary}, and is 
the largest one could reasonably expect in a heavy ion collision.


\subsubsection{How this analysis can inform the experimental 
search for the critical point }
\label{sec:exp}

We have analyzed the relevant length scales for the critical point search. In 
heavy ion collisions the longest wavelengths are long range in rapidity,
and are described with hydrodynamics. These
long wavelength modes, such as the elliptic and triangular flow,  
are not equilibrated and depend on the initial conditions. 
Only wavelengths smaller than a characteristic scale $\ell_{\rm max}$ 
equilibrate  during an expansion away from the critical point.  Only modes with
$({\rm wavelength}) \ll \ell_{\rm max}$
can possibly exhibit critical properties.  The 
typical wavelength for enhanced critical fluctuations is set by 
the Kibble-Zurek length $\ell_{\rm kz}$, and this length is only somewhat larger
than the inter-particle spacing in practice. Such short lengths are associated with non-flow correlations.
Thus, if critical fluctuations  are to be seen then one  must carefully examine
the non-flow correlations to look for modifications as the mean baryon
number to entropy ratio is changed in the event.

The current measurements of kurtosis are essentially a measure
of the probability of finding a baryon at mid-rapidity while keeping
the particle number (entropy) fixed. It seems to us that the modifications of
this quantity with beam energy are  mostly a measurement of baryon transport in the initial state, and are perhaps unrelated to the critical fluctuations.

In order to measure the expected critical point signal, one should divide the
system at different beam energies into different event classes with a specified
$n/s$ in a large mid-rapidity detector. (The proton to pion ratio can be used
as a proxy for $n/s$.)   If the system passes close to the critical point, the
short range (connected) two point functions should change as the $n/s$ event
class is scanned. These changes in the two point functions should be largely
independent of centrality and beam energy, but should depend only on the mean $n/s$
of the event class. 
The presence of a critical point leads to short range spatial correlations of size of order $\ell_{\rm kz}$. In momentum space this corresponds to a momentum difference of order $\Delta p \sim \hbar/{\ell_{\rm kz}} \sim 50\, {\rm MeV}$.  
Thus, the presence of a critical point there will enhance the short-range, almost HBT-like, correlations.

Any non-monotonic changes in the non-flow correlation strength  in this
fixed momentum range with the mean $n/s$
 would certainly be remarkable. We plan to investigate such
correlations in future work, and encourage our experimental colleagues to do the
same.

\begin{acknowledgments}
We thank Aleksas Mazeliauskas for collaboration during the initial stages of this project.  
We are grateful to Jiunn-wei Chen, Prithwish Tribedy, Xiaofeng Luo,
Misha Stephanov for helpful conversations.
This work is supported by JSPS KAKENHI Grant Number JP18K13538 (Y.A.) and by the U.S. Department of Energy, Office of Science, Office of Nuclear Physics, within the framework of the Beam Energy Scan Theory (BEST) Topical Collaboration (Y.Y.) and grants Nos. DE\nobreakdash-FG\nobreakdash-02\nobreakdash-08ER41450 (D.T. , F.Y) and DE-SC0011090 (Y.Y)
\end{acknowledgments}

\clearpage

\begin{appendices}

\section{The Ising equation of state  and correlation length} \label{appendix:eos}

In this section we will parametrize the Ising  equation of state  with the familiar $(R,\theta)$ form.

\subsubsection{Preliminaries}
The free energy is the log of the partition function\footnote{ 
Relative to \Ref{Nonaka:2004pg}, but in accord with  \Ref{onuki2002phase},
we have reversed the roles of $F$ (what we call the free energy) and $G$ (what we call the Gibbs free energy) }
\st
   F(T,H) = -T  \frac{\log Z(T, H)}{V}\, ,  \qquad \mbox{with} \qquad dF =  - S dT   -  \psi dH \, ,
\stp
and thus 
\st
  \label{eq:dZtotal}
  \frac{d \log Z(T,H)}{V} =  \frac{\mathcal E }{T^2} dT + \frac{\psi}{T} dH \, , 
\stp
where the energy density is $\mathcal E = F - T \frac{\partial F}{\partial T}$.  Near the critical point 
$Z(T,H)$ is the product of a regular contribution and a singular 
contribution, $Z_{\rm reg} \times Z_{\rm sing}$. The regular part
is expanded in a Taylor series near the critical point, keeping only linear terms
\st
 \label{eq:dZreg}
 \frac{\Delta \log Z_{\rm reg}}{V} = \mathcal E_c  \frac{\Delta T}{T_c^2} = -\frac{\Delta F_{\rm reg} }{T_c} \, ,  \qquad  \Delta F_{\rm reg} =   S_c  \Delta T \, .
\stp
 Due to the $Z_2$ symmetry of the Ising model, the regular part starts as $H^2$ which can be neglected close
to the critical point.
Given \Eq{eq:dZtotal} and \Eq{eq:dZreg} the  singular contribution  $Z_{\rm sing}$  satisfies
\st
 d\log Z_{\rm sing} = -\frac{dF_{\rm sing}}{T_c} =\epsilon \, dr +  \psi \, dh\,,    \qquad  \frac{dF_{\rm sing}}{T_c} =  -s \, dr - \psi \,dh \,,
\stp
where we have defined $r = (T - T_c)/T_c$ , $\epsilon = (\mathcal E - \mathcal E_c)/T_c$, $h = H/T_c$, and $s=S(T,H) - S_c$.  Thus, near the critical point we
have $\epsilon=s$ which follows from the definition of $\mathcal E$, 
$\epsilon$, and the decomposition of  $Z=Z_{\rm reg} \times Z_{\rm sing}$ into regular and singular parts.

The free energy $F$ is the Legendre transform of the Gibbs free energy $G(T,\psi) = F + \psi H$. The singular  part satisfies 
\st
\label{logZsing}
   \log Z_{\rm sing}(r,h)  =  -\frac{G_{\rm sing}(r,\psi)}{T_c}  + \psi h \, ,
\stp
and the reduced magnetic field $h$ is related to $G_{\rm sing}(r,\psi)/T_c$ by the thermodynamic relations,  $h = \left(\partial (G_{\rm sing}/T_c)/\partial \psi\right)_r$.  

\subsubsection{The $(R,\theta)$ parameterization }

Following  previous authors~\cite{onuki2002phase,Nonaka:2004pg}, we 
parametrize the Ising equation of state outside of the coexistence region with
two auxiliary variables $(R, \theta)$  with $\theta^2 \leq \theta_0^2$
\begin{align}
   \label{Rthetadef}
   \ris =& (1 - \theta^2 ) R\,,  \\
   \frac{h}{h_0} =& c_h \, \theta\left(1- \frac{\theta^2}{\theta_0^2}\right) R^{\beta \delta}\, .
\end{align}
Then the equation of state takes the form~\cite{onuki2002phase}
\begin{align}
   \label{EOS1}
   \frac{\psi}{\M_0} =&  c_{\M}  \, \theta R^{\beta} \, ,
\end{align}
where $\delta$ and $\beta$ are critical exponents. $\theta_0$ demarcates 
the boundary of the coexistence region and is approximately\footnote{
   The differences between our parameterization (taken from  \Ref{onuki2002phase}) and the parametrization used in \Ref{Nonaka:2004pg} are
   minor. We have neglected the fifth order term in 
   the polynomial expansion of 
   $\tilde{h}(\theta)\simeq \theta (1 - \theta^2/\theta_0^2) $, and taken
   an analytic expression  (valid to $\epsilon^2$ in the $\epsilon$ expansion) for the first zero $\theta_0$ of $\tilde{h}(\theta)$~\cite{onuki2002phase}.  
   With this simplified parametrization the specific heat $C_{M}$ is only a function of $R$ and the susceptibilities take a compact form.
   The numerical accuracy of this parametrization is more than sufficient for heavy ion physics. 
}
\begin{align}
\label{theta}
      \theta_0 = \left(\frac{\delta-3}{(\delta -1)(1 - 2\beta) } \right)^{1/2} \simeq 1.166 \,.
\end{align}

As discussed in \Sect{sec:map}, the dimensionful constants $M_0 h_0$ and $\mathcal M_0$ are chosen conventionally to be $(n_c, s_c)$ so that mapping matrix $M^{A}_{\;\;b}$ is 
of order unity.
  The constants $c_h$ and $c_{M}$ will be chosen to maintain the convenient
  normalization conventions adopted in \Sect{sec:timescale}:  namely that on coexistence
  line $\psi/\M_0= |r|^\beta$ and  $\epsilon/(\M_0 h_0)=-|r|^{1-\alpha}$. Thus 
\st
\label{eq:cm}
c_\M = \frac{(\theta_0^2-1)^\beta}{\theta_0} \simeq 0.6145 \, ,
\stp
and $c_\M c_h$ is given below in \Eq{eq:chcm}.

The dimensionless scaling variable $\theta$ is directly related
to the scaling variable used in\footnote{
   Here our $(h/c_h h_0)$  and $h_S$ 
   are denoted by $H$ and $H_0$ respectively by \Ref{Engels:2002fi}
} \Ref{Engels:2002fi}
\st
   \label{zdef}
   z =  \left(\frac{\ris}{\ris_S}\right) \left( \frac{ h_S}{h/(c_h h_0)} \right)^{1/\beta\delta}=1.901 \frac{(1-\theta^2) }{\left[\theta(1- (\theta/\theta_0)^2)\right]^{1/\beta\delta}} \, ,
\stp
where we defined 
\st
\ris_S = \frac{ \theta_0^2-1}{\theta_0^{1/\beta}}\simeq 0.225\,, \quad \mbox{and} \quad h_S = \frac{\theta_0^2 - 1}{\theta_0^2}=0.265. 
\stp

Following \Ref{Nonaka:2004pg},  we can integrate the equation of state,
\Eq{EOS1}, to determine the singular part of the grand sum,  $\log Z(r, h)$, which subsequently determines all thermodynamic quantities
and susceptibilities through differentiation.
Parametrizing
  $G_{\rm sing}/T_c$  as
\st
\label{gsing}
\frac{1}{h_0 \M_0} \frac{G_{\rm sing} (r,\psi)}{T_c} = c_h c_\M R^{2- \alpha} g(\theta) \, ,
\stp
a differential equation is easily obtained for $g(\theta)$:
\st
(1- \theta ^2) g'(\theta )+2 (2-\alpha) \theta  g(\theta ) = 
\left( 2 \beta\theta^2 + (1 - \theta ^2) \right) \theta (1- (\theta/\theta_0)^2)^2 ) \, .
\stp
Integrating the differential equation we find
\st
g(\theta) =  \frac{(2 \beta -1) \left(\theta ^2-1\right)^2}{2 \alpha  \theta _0^2}+\frac{\left(\theta
   ^2-1\right) \left((1-2 \beta ) \theta _0^2+4 \beta -1\right)}{2 (\alpha -1) \theta
   _0^2}-\frac{\beta  \left(\theta _0^2-1\right)}{(\alpha -2) \theta _0^2} \, , 
\stp
up to a homogeneous solution which does not contribute to the singular behavior~\cite{Nonaka:2004pg}.  

From these expressions first derivatives can be obtained
\begin{align}
\begin{pmatrix}
-s & h  
\end{pmatrix}
=
   \frac{\partial (G_{\rm sing}/T_c)}{\partial(r, \psi)  }
   =&  \frac{\partial (G_{\rm sing}/T_c)}{\partial(R, \theta) } 
\left( \frac{\partial (R, \theta)  }{\partial (r, \psi) } \right) \, ,
\end{align}
where  in practice this Jacobian matrix is evaluated through
its inverse
\st
\left( \frac{\partial (R, \theta)  }{\partial (r, \psi) } \right) 
=  \left( \frac{\partial (r , \psi)  }{\partial (R,\theta) } \right)^{-1} \, .
\stp
The singular entropy density and the singular energy density take the form
\st
\frac{\epsilon}{ \M_0 h_0} = \frac{s}{\M_0 h_0}  =  c_\M  c_h \, R^{1-\alpha} f_{\eis}(\theta)
\stp
with
\begin{align}
f_{\eis}(\theta) =& \frac{\beta (1-\delta)  \left(-(1-\alpha) (2 \beta-1)
    \theta^2+\alpha+2 \beta-1\right)}{2 (1-\alpha) \alpha} \,, \\
\simeq & 
1.496-1.951 \theta^2 \, .
\end{align}
From our requirement that on the coexistence line that  $\epsilon/(\M_0 h_0) = -|r|^{1-\alpha}$ 
we find
\st
\label{eq:chcm}
c_{\M} c_h =  -\frac{(\theta_0^2-1)^{1-\alpha} }{f_\epsilon(\theta_0) } \simeq  0.3486 \, .
\stp

In a similar way the susceptibility matrix 
can be computed by taking second derivatives of the partition function, yielding
\begin{subequations}
   \label{eq:susceptibility_explicit}
\begin{align}
   \frac{ C_{M}}{\M_0 h_0 }  =& c_\M c_h  \, \frac{\gamma(\gamma-1)}{2\alpha} \, R^{-\alpha}  \, , \\
   \frac{ \chi}{(\M_0/h_0)}  =& \frac{c_\M}{c_h } \left[1 + (2\beta\delta - 3)(\delta-1)\theta^2/(\delta-3)\right]^{-1} R^{-\gamma}\, , \label{chieqnapp} \\
   \frac{C_{H}}{\M_0 h_0 }  =& c_\M  c_h \,   \frac{\gamma}{2\alpha} \left[ \frac{(2 \beta-1) (\delta-1) (\beta (\delta+3)-3)\theta^2 +(\delta-3) (\gamma-1)
   }{(\delta-1) (2 \beta \delta -3) \theta^2+(\delta-3)} \right] R^{-\alpha} \, .
\end{align}
\end{subequations}
It is particularly noteworthy that $C_{M}$ is independent of the angle $\theta$.

\subsubsection{The correlation length}
\label{sec:correlation}

To evaluate the correlation length we used the numerical data from 
Engels, Fromme and Seniuch (EFS)~\cite{Engels:2002fi} which is expressed in terms of the scaling variable
$z$ given in \Eq{zdef}.  The correlation length takes the scaling form
\st
\label{xihandz}
\xi(h,z) = \left(\frac{h/(c_h h_0)}{h_S}\right)^{-\nu/\beta\delta} g_{\xi}(z)   \, .
\stp
where $g_{\xi}(z)$ is a universal function (up to its normalization), which was determined numerically
through precise simulations of the Ising model. Even its normalization
is not independent of the non-universal parameters, $\M_0$ and $h_0$, introduced previously.

Since $g_\xi(z) \propto z^{-\nu}$ for $z$ large, the correlation length
at zero field and $T > T_c$ behaves as
\st
 \xi \xrightarrow[z\rightarrow \infty]{} \xi_+ r^{-\nu} \, ,
\stp
where we have used the definition of $z$ given in \Eq{zdef}.
The length scale $\xi_+$ is not independent of $\mathcal M_0$ and $h_0$,
but is fixed from the scaling  form the free energy
\st
  -\frac{ F_{\rm sing}}{T_c} = \frac{\log Z_{\rm sing}(r, h) }{V}= \xi^{-d} \mathcal F_{\rm sing}(z) \, , 
\stp 
where $d=3$ notates the number of spatial dimensions, and $\mathcal F_{\rm sing}(z)$ is a universal function.
Comparison with \Eq{gsing} suggests that $\mathcal M_0 h_0 (\xi_+)^d$ should be a universal
constant~\cite{onuki2002phase}. Indeed, EFS  relate $\xi_+$ to the parameters of the equation of state, $\mathcal M_0$ and $h_0$, introduced above. Translating their ratio
into the current notation we have\footnote{They define the
parameters $B$ and $C_+$  which 
 in the current notation  read:
\st
B =  \frac{\theta_0}{ (\theta_0^2 - 1)^\beta} \, (c_\M \M_0) \simeq 1.6274\, (c_\M {\mathcal M}_0)\,, \qquad  C_+= \frac{c_\M \mathcal M_0}{c_h h_0} \, .
\stp
They numerically determine the amplitude ratio 
$Q_c = B^2 (\xi_+)^d / C_+ = 0.326$ 
which determines \Eq{normalizedplus}. 
}
\st
\label{normalizedplus}
(c_\M c_h) \, \mathcal M_0 h_0 \, \xi_+^d  = 0.1231\, .
\stp

In  EFS, the numerical data for a normalized $g_{\xi}(z)$ 
is presented by comparing it to  the scaling function of the susceptibility.
Specifically, the susceptibility $\chi$ (see \Eq{chieqnapp}) is written
\st
\chi =  \frac{1}{h_S} \left(\frac{h/(c_h h_0)}{h_S}\right)^{1/\delta -1} f_{\chi}(z) \, , 
\stp
where $h_S$ and its relation to the notation of EFS is given  in \Eq{zdef} and the corresponding footnote. $f_{\chi}(z)$ has
the asymptotic form
\begin{align}
   f_{\chi} \xrightarrow[ z\rightarrow +\infty ]{}  R_{\chi} z^{-\gamma} \,,
\end{align}
with  $R_\chi\simeq1.723$.
$g_{\xi}(z)$ is normalized and scaled by $f_{\chi}(z)$
\st
\label{gxiparam}
    g_{\xi}(z) = g_{\xi}(0) \left(f_{\chi}(z)\right)^{1/2} \left( \frac{\hat g_{\xi}^2(z) }{f_{\chi}(z)} \right)^{1/2} \, .
\stp
We fit the numerical data in Fig.~11 of EFS with 
\st
\label{fitform}
\frac{\hat g^2_{\xi}(z) }{f_\chi(z)}= 
\frac{ (u_+  + u_{-}) -  (u_- - u_+) \tanh( (z - x_0)/\sigma) }{2 ((z - x_0)^2 + 1)^{\eta\nu/2}  } \, ,
\stp
which has the correct asymptotics
\st
\frac{\hat g^2_{\xi}(z) }{f_\chi(z)}  \xrightarrow[z \rightarrow \pm \infty]{}   u_{\pm} z^{-\eta \nu/2} \, .
\stp
The parameters are $u_+$, $u_-$, and $\sigma$ from the fit are
\st
   u_+ =4.133\,, \qquad  u_{-} = 5.32\,,   \qquad \sigma=3\,,  \qquad x_0 = 0.3431\,.
\stp
The value $x_0=0.3431$ is constrained by the universality requirement 
that  at $z=0$  we have $\hat g^2_\xi/f_{\chi} = \delta$.
The slight deviation in our fitted values of $u_{+}$ and $u_{-}$ 
from the asymptotic values  quoted by EFS ($u_+=4.001$ 
and $u_+/u_-\simeq 0.75$ respectively) stems from a desire to have a somewhat better fit
over the full range in $z$.
Finally,  with the functional form given in \Eq{fitform} and
the normalization in \Eq{normalizedplus}, the value of 
$g_{\xi}(0)$ of zero can be determined
\st
\label{gxizero}
  g_\xi(0) = \frac{0.4838 }{(c_\M c_h \, \mathcal M_0 h_0)^{1/d}} \,, 
\stp
where we have unraveled the nested definitions to
establish that $\xi_{+} =  g_{\xi}(0) r_S^{\nu} (u_+  R_\chi)^{1/2}$. 

Summarizing, we use Eqs.~(\ref{xihandz}), (\ref{gxiparam}),(\ref{fitform}), and (\ref{gxizero}) to evaluate the correlation length for any given value 
of $h,r$.

%






\end{appendices}

\bibliography{refs}

\end{document}